\documentclass[%
 reprint,
superscriptaddress,
 amsmath,amssymb,
 aps,
floatfix,
]{revtex4-2}

\usepackage{graphicx}
\usepackage{dcolumn}
\usepackage{bm}
\usepackage{hyperref}

\usepackage[utf8]{inputenc}
\usepackage[T1]{fontenc}
\usepackage{mathptmx}
\usepackage{gensymb}

\DeclareUnicodeCharacter{2212}{-}

\begin{document}

\preprint{APS/123-QED}

\title{The connection between polymer collapse and the onset of jamming}

\author{Alex T. Grigas}
\affiliation{Graduate Program in Computational Biology and Bioinformatics, Yale University, New Haven, Connecticut, 06520, USA}
\affiliation{Integrated Graduate Program in Physical and Engineering Biology, Yale University, New Haven, Connecticut, 06520, USA}
\author{Aliza Fisher}
\affiliation{Department of Mechanical Engineering and Materials Science, Yale University, New Haven, Connecticut, 06520, USA}
\author{Mark D. Shattuck}
\affiliation{Benjamin Levich Institute and Physics Department,
The City College of New York, New York, New York 10031, USA}
\author{Corey S. O'Hern}
\affiliation{Department of Mechanical Engineering and Materials Science, Yale University, New Haven, Connecticut, 06520, USA}
\affiliation{Graduate Program in Computational Biology and Bioinformatics, Yale University, New Haven, Connecticut, 06520, USA}
\affiliation{Integrated Graduate Program in Physical and Engineering Biology, Yale University, New Haven, Connecticut, 06520, USA}
\affiliation{Department of Physics, Yale University, New Haven, Connecticut, 06520, USA}
\affiliation{Department of Applied Physics, Yale University, New Haven, Connecticut, 06520, USA}

\date{\today}

\begin{abstract}
Previous studies have shown that the interiors of proteins are densely packed, reaching packing fractions that are as large as those found for static packings of individual amino-acid-shaped particles. How can the interiors of proteins take on such high packing fractions given that amino acids are connected by peptide bonds and many amino acids are hydrophobic with attractive interactions? We investigate this question by comparing the structural and mechanical properties of collapsed attractive disk-shaped bead-spring polymers to those of three reference systems: static packings of repulsive disks, of attractive disks, and of repulsive disk-shaped bead-spring polymers. We show that attractive systems quenched to temperatures below the glass transition $T \ll T_g$ and static packings of both repulsive disks and bead-spring polymers possess similar interior packing fractions. Previous studies have shown that static packings of repulsive disks are isostatic at jamming onset, i.e. the number of contacts $N_c$ matches the number of degrees of freedom, which strongly influences their mechanical properties. We find that repulsive polymers are hypostatic at jamming onset, but effectively isostatic when including quartic modes. While attractive disk and polymer packings are hyperstatic, we identify a definition for interparticle contacts for which they can also be considered as effectively isostatic. As a result, we show that the mechanical properties (e.g. scaling of the potential energy with excess contact number and low-frequency contribution to the density of vibrational modes) of weakly attractive disk and polymer packings are similar to those of {\it isostatic} repulsive disk and polymer packings. Our results demonstrate that static packings generated via attractive collapse or compression of repulsive particles possess similar structural and mechanical properties.
\end{abstract}

\maketitle


\section{\label{sec:intro} Introduction}

It has long been appreciated since the first atomic-resolution x-ray crystal structures of proteins were solved that their interior, solvent inaccessible, or core, regions are densely packed, regardless of the differences in their overall folds~\cite{packing:RichardsJMB1974,packing:ChothiaNature1975,packing:RichardsAnnRevBiophys1977,packing:TsaiJMB1999,packing:LiangBPJ2001,subgroup:GainesPRE2016}. Other experimental atomic-scale structural characterization methods, such as NMR spectroscopy, provide all-atom structures of proteins in solution and at room temperature, and have shown that high-quality NMR structures also possess densely packed interiors with packing fractions similar to those of x-ray crystal structures~\cite{subgroup:GrigasProSci2022}. Additionally, perturbing the dense packing of the solvent-inaccessible hydrophobic interior of proteins via mutation has been shown to significantly affect protein structure and stability~\cite{folding:DillBiochemistry1990,stability:XuProtSci1998,stability:BaaseProtSci2010,folding:PaceJMB2011}.

Prior analyses of protein x-ray crystal structures that allowed unphysical atomic overlaps suggested that the interiors of proteins possessed packing fractions as large as $\phi \sim 0.7-0.75$~\cite{packing:RichardsJMB1974,packing:LiangBPJ2001}. However, more recent studies that account for the non-spherical shapes of amino acids and do not allow atomic overlaps have shown that the average packing fraction of solvent inaccessible amino acids is $\phi \approx 0.55 \pm 0.02$~\cite{subgroup:GainesPRE2016,subgroup:GainesJPhysConMat2017,subgroup:GainesProteins2018,subgroup:TreadoPRE2019}. Why do the core regions of all experimentally determined protein structures, regardless of the overall fold, possess this value for the packing fraction? Previously, we have shown that jammed packings of rigid amino-acid-shaped particles with purely repulsive interactions under periodic boundary conditions possess similar packing fraction distributions as those for experimentally determined protein cores~\cite{subgroup:GainesPRE2016}. Despite this agreement, these prior simulations lacked important features of protein structure: the amino acids were rigid with no backbone dihedral angle degrees of freedom and they were {\it disconnected}, lacking peptide bonds; the packings were generated by compression, not by hydrophobic polymer collapse; and the packings were generated using periodic boundary conditions instead of being fully solvated. In addition, when thermal fluctuations are included in the amino-acid-shaped particle-packing generation protocol, we find that the onset of jamming occurs over a range of packing fractions, $0.55 \lesssim \phi_J \lesssim 0.62$, where $\phi_J$ increases as the rate at which thermal energy is removed from the system decreases~\cite{subgroup:MeiProteins2020,packing:SeeligerProteins2007}. To date, the only high-resolution experimentally determined protein cores that possess $\phi \gtrsim 0.55$ were solved using x-ray crystallography at extremely high pressures~\cite{highPxtal:YamadaActaCrystD2015}. Does the correspondence between the packing fraction of jammed packings of repulsive, disconnected amino-acid-shaped particles generated via rapid compression and the cores of experimentally determined proteins indicate a deep connection between the two systems or is it fortuitous? 

More generally, to isolate the essential features of the problem, we can ask, for connected and disconnected spherical particles, what is the relationship between the thermal collapse of sticky, spherical bead-spring polymers or aggregation of sticky spherical particles and the onset of jamming of purely repulsive spherical particles under athermal, quasi-static compression? Here, we focus specifically on disk-shaped particles versus disk-shaped bead-spring polymers and purely repulsive potentials versus potentials with both short-range repulsive and longer-range attractive interactions in two dimensions (2D). 

Mechanically stable (or jammed) packings of repulsive spherical particles are isostatic, i.e. the number of constraints arising from interparticle and particle-boundary contacts matches the number of degrees of freedom, which strongly influences their structural and mechanical properties~\cite{jamming:OHernPRE2003}. Prior studies have shown that isostatic sphere packings at jamming onset can occur over a range of packing fractions (known as the J-line), from a lower bound similar to values quoted for random close packing and increasing as the compression rate and rate of energy relaxation decrease~\cite{jamming:ChaudhuriPRL2010,jamming:AshwinPRL2013,polydisperse:OzawaSciPost2017}. Isostatic jammed sphere packings also possess an excess low frequency contribution to the vibrational density of states $D(\omega)$, which is quantified by a characteristic frequency $\omega^{\ast}$ that increases as the packings are compressed above jamming onset. Further, the shear modulus and $\omega^{\ast}$ obey power-law scaling relations with the deviation $\Delta z$ of the coordination number from that at jamming onset. 

Previous work has also suggested that repulsive spherical bead-spring polymers compressed to jamming onset are nearly isostatic even though they possess fixed constraints through the polymer backbone~\cite{packing:KarayiannisPRL2008,packing:KarayiannisJCP2009,polymerpacking:LopatinaPRE2011,polymerpacking:HoyPRL2017,packing:SoikPRE2019}. As found for jammed sphere packings, jammed repulsive polymer packings occur over a range of packing fractions when they are generated using different protocols, but it is unclear whether this range of packing fractions is the same as that for jammed sphere packings. Further, it has been suggested that the elastic moduli of jammed repulsive polymer packings are similar to those of jammed sphere packings~\cite{packing:SoikPRE2019}.

Collections of spherical monomers with attractive interactions are generally not isostatic. For example, attractive, spherical particles can form sparse, yet solid-like gels at extremely low packing fractions with on average two contacts per particle. They can also form dense, attractive glasses, where each particle possesses close to an isostatic number of nearest-neighbor contacts and many more longer-range interactions~\cite{sticky:LoisPRL2008,sticky:KoezePRL2018,sticky:KoezePRR2020}. Spherical bead-spring polymers with attractive interactions collapse into dense liquid globules at sufficiently low temperatures~\cite{polymercollapse:WilliamsAnnRev1981}.  Further decreasing the temperature will generate collapsed {\it glassy} globules with a wide range of structural and mechanical properties~\cite{sticky:PaulJCP2001,polymerpacking:HoyPRL2010}. Despite this fact, we have found in previous studies that the interiors of folded proteins (that possess both short-range repulsive and longer-range attractive interactions) appear to share properties with jammed packings of disconnected, repulsive amino-acid-shaped particles generated via athermal, quasi-static compression.

Here, to understand the connection between the thermal collapse of sticky polymers and jamming of repulsive particles under athermal compression, we compare the interior packing fractions of static packings of single disk-shaped bead-spring polymers and static packings of disconnected disks, with either attractive or repulsive interactions, as shown in Fig.~\ref{fig:system}. For systems with non-bonded attractive interactions, we study the interior packing fraction as the system is cooled below the glass transition temperature at varying rates. For systems with purely repulsive non-bonded interactions, we develop an open-boundary ``jamming'' protocol where the system undergoes athermal, quasi-static compression until reaching a mechanically stable state using an externally applied radial force. 

We find several important results. First, for a collapsed polymer with attractive non-bonded interactions to obtain interior packing fractions $\phi$ similar to those found for jammed packings of purely repulsive disks, they must be quenched well below the glass transition temperature. Additionally, we find that the attractive systems (both monomeric and polymeric) quenched to zero temperature and the repulsive systems (both disks and polymes) compressed to jamming onset with open boundary conditions possess similar interior packing fractions for all system sizes, damping parameters, and initial temperatures studied. We show that packings of attractive disks and polymers possess excess low-frequency vibrational modes in the limit of small attractive strength. As expected, we find that repulsive disks compressed to jamming onset are isostatic. In contrast to prior work, we find that packings of polymers with non-bonded repulsive interactions are hypostatic at jamming onset. However, the number of missing contacts matches the number of quartic modes, and thus packings of repulsive polymers are effectively isostatic. While packings of attractive monomers and polymers are hyperstatic when counting contacts using the full interaction potential, they can also be considered to be effectively isostatic if we appropriately re-define the interparticle contact network. By varying the attractive strength, we observe the same scaling of the low-frequency modes of $D(\omega)$ and excess number of contacts $\Delta N$ from the isostatic number versus the potential energy as found for repulsive disk packings compressed above jamming onset.

This article is organized into three additional sections and two appendices. In Sec.~\ref{sec:methods}, we describe the numerical models for the disk-shaped bead-spring polymers and disk-shaped monomers with non-bonded attractive and repulsive interactions, the packing generation protocols, and how we identify surface versus core disks for the calculation of the interior packing fraction. In Sec.~\ref{sec:results}, we present the results for the interior packing fraction, characteristic plateau frequency of the distribution of vibrational modes $D(\omega)$, and contact number for each system. Finally, in Section~\ref{sec:conclusions}, we discuss the implications of the results for understanding the dynamics of polymer collapse and protein folding and propose future work on athermal compression of all-atom models of proteins to jamming onset. In Appendix~\ref{sec:jamming_appendix}, we describe methods to avoid size segregation when applying a radial force to generate jammed packings of repulsive monomers and polymers in open boundary conditions and in Appendix~\ref{sec:rsasa_appendix}, we provide additional details of the algorithm for identifying interior versus surface particles.

\section{\label{sec:methods}Methods}

\subsection{Model systems}
\label{model}

\begin{figure}[h!]
\begin{center}
\includegraphics[width=0.875\columnwidth]{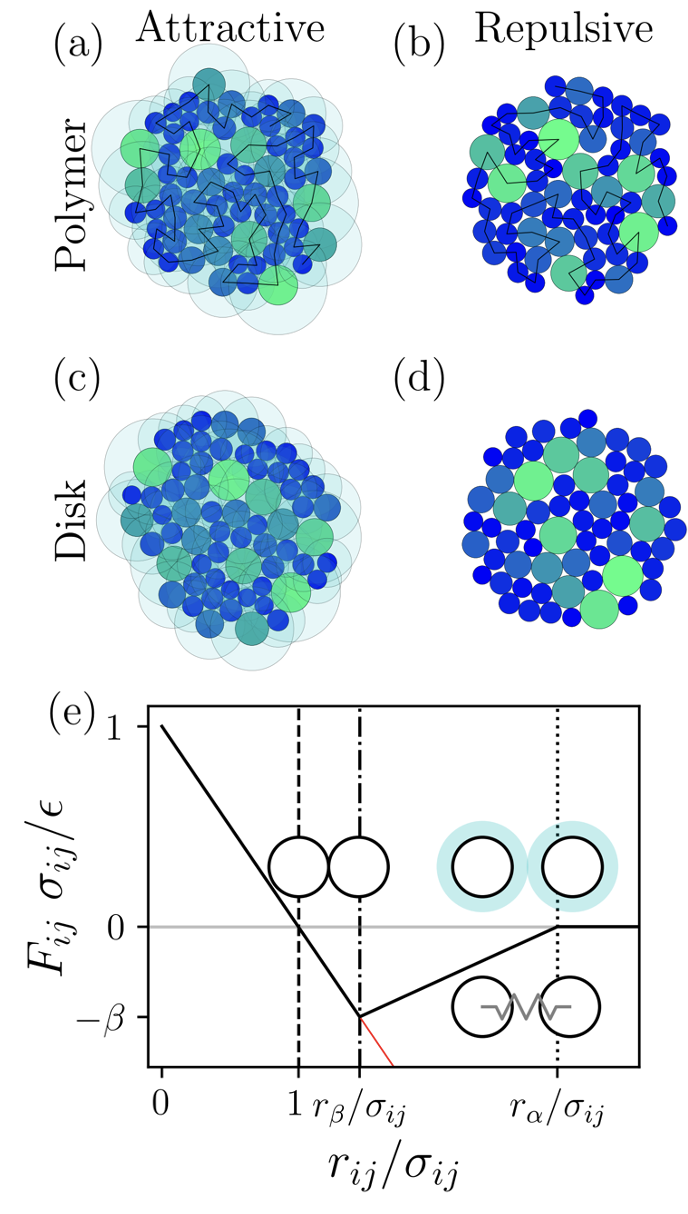}
\caption{Example static packings for a single disk-shaped bead-spring polymer (a) with and (b) without attractive interactions and disk-shaped monomers (c) with and (d) without attractive interactions. The disk diameters are polydisperse, obeying an inverse power-law distribution for the diameters; the color shading indicates the particle size from large to small (light green to blue). The cyan shading in (a) and (c) indicates the range of the attractive interactions with $\alpha=1.5$ (Eq.~\ref{eq:ra}). The black solid lines connecting adjacent disks indicate the polymer backbone. (e) Force magnitude $F_{ij} \sigma_{ij}/\epsilon$ between disks $i$ and $j$ plotted versus their separation $r_{ij}$ normalized by their average diameter $\sigma_{ij} = (\sigma_i + \sigma_j) / 2$. For repulsive non-bonded interactions, the disks interact only when they overlap and are repelled by a repulsive linear spring force for $r_{ij} < \sigma_{ij}$ (vertical black dashed line). Repulsive polymers include the same repulsive interactions and extend the interaction for $r_{ij} > \sigma_{ij}$ to a double-sided linear spring for bonded disks (red thin solid line). Non-bonded attractive interactions are specified by an attractive range $\alpha$ and strength $\beta$; in this case, the non-bonded force is extended to $F_{ij} (r_{\beta}/\sigma_{ij}) \sigma_{ij}/\epsilon=  -\beta$, where $r_{\beta}/\sigma_{ij} = 1 + \beta$ (vertical red dot-dashed line), after which the force linearly returns to zero at $r_{\alpha}/\sigma_{ij} = 1+\alpha$ (vertical grey dotted line).}
\label{fig:system}
\end{center}
\end{figure}

We study four types of systems: single disk-shaped bead-spring polymers with attractive non-bonded interactions, attractive disks (or monomers), single disk-shaped bead-spring polymers with repulsive non-bonded interactions, and repulsive disks (or monomers) as shown in Fig.~\ref{fig:system}. The non-bonded, repulsive interactions are modeled by the repulsive linear spring potential,
\begin{equation}
\label{eq:repulsive}
    \frac{V^{rnb}(r_{ij})}{\epsilon} = \frac{1}{2} \left( 1 - \frac{r_{ij}}{\sigma_{ij}} \right)^2 \Theta \left(1-\frac{r_{ij}}{\sigma_{ij}} \right),
\end{equation}
where $r_{ij}$ is the center-to-center distance between disks $i$ and $j$, $\sigma_{ij}$ is their average diameter, $\epsilon$ is the energy scale of the repulsive interaction, and $\Theta \left( x \right)$ is the Heaviside step-function. 
For the $N-1$ bonded interactions between disks $i$ and $j=i+1$ in the bead-spring polymer, the repulsive linear spring potential is extended into a double-sided linear spring potential: 
\begin{equation}
\label{eq:bond}
    \frac{V^{b}(r_{ij})}{\epsilon} = \frac{1}{2} \left( 1 - \frac{r_{ij}}{\sigma_{ij}} \right)^2 . 
\end{equation}
We parameterize the non-bonded attractive interactions by the attractive cutoff distance $\alpha$ and depth $\beta$. Previous work on jamming of spherical particles with short-ranged attractive interactions used a single parameter to characterize the attractive interactions~\cite{sticky:LoisPRL2008,sticky:KoezePRL2018,sticky:KoezePRR2020}. Here, we separate the attractive range and depth to allow the model to capture both short-ranged, sticky disks and molecular liquids with weak, but long-range attractive interactions. For the non-bonded attractive interactions, we extend the potential in Eq.~\ref{eq:repulsive} to $r_{\beta} > \sigma_{ij}$ and cutoff the interactions at $r_{\alpha}= (1+\alpha)\sigma_{ij} > r_{\beta}$:
\begin{equation}
\label{eq:ra}
\frac{V^{anb}(r_{ij})}{\epsilon} = \begin{cases}
    \frac{1}{2} \left( 1 - \frac{r_{ij}}{\sigma_{ij}} \right)^2 - V_c/\epsilon ~~{\rm for}~r_{ij} \leq r_{\beta} \\
    -\frac{k}{2\epsilon} \left( \frac{r_{ij}}{r_{\alpha}} - 1 \right)^2 ~~~~~{\rm for}~ r_{\beta} < r_{ij} \leq r_{\alpha} \\
    0 ~~~~~~~~~~~~~~~~~~~~~~~~~~~~{\rm for}~ r_{ij} > r_{\alpha},
\end{cases}
\end{equation}
where $V_c/\epsilon =  (k/\epsilon)\left( r_{\beta}/r_{\alpha} - 1 \right)^2/2  + \left( 1 - r_{\beta}/\sigma_{ij} \right)^2/2$. The pair potential energy for attractive polymers (Fig.~\ref{fig:system} (a)) is $V(r_{ij}) = V^{b}(r_{ij}) + V^{anb}(r_{ij})$. For repulsive polymers (Fig.~\ref{fig:system} (b)), $V(r_{ij}) = V^{b}(r_{ij}) + V^{rnb}(r_{ij})$. For attractive disks (Fig.~\ref{fig:system} (c)), $V(r_{ij}) = V^{anb}(r_{ij})$ and for repulsive disks, $V(r_{ij}) = V^{rnb}(r_{ij})$ (Fig.~\ref{fig:system} (d)). The total potential energy and interparticle forces for each system are given by $V= \sum_{i > j} V(r_{ij})$ and ${\vec F}_{ij} = -(dV/dr_{ij}){\hat r}_{ij}$.  Note that we set $F_{ij}(r_{\beta})=-\epsilon \beta/\sigma_{ij}$ and $k/\epsilon= (\beta r_{\alpha}/\sigma_{ij}) \left(r_{\beta}/r_{\alpha} - 1 \right)$ to  
ensure that the non-bonded forces are continuous as shown in Fig.~\ref{fig:system} (e). Below, we consider dimensionless forces $F_{ij} \sigma_s/\epsilon$, potential energies $V/\epsilon$, frequencies $\sqrt{\epsilon/m}\sigma_s^{-1}$, and temperature $k_b T/\epsilon$, where $k_b=1$ is the Boltzmann constant, $m$ is the mass of each disk, and $\sigma_s$ is the size of the smallest disk.

To prevent crystallization~\cite{polydisperse:OzawaSciPost2017} during the packing generation process, the disk diameters are selected randomly from a power-law size distribution, $P(\sigma_i) = A \sigma_i^{-3}$, with minimum and maximum diameters $\sigma_s$ and $\sigma_{\rm max} = 2.2 \sigma_s$ and polydispersity $D = (\langle \sigma_i^2\rangle - \langle \sigma_i \rangle^2)/\langle \sigma_i \rangle^2 \sim 0.23$. For each system size of $N$ disks, we average over $100$ different sets of diameters $\{ \sigma_{i} \}$ that were randomly selected from $P(\sigma_i)$.

\subsection{Packing-generation protocol}

Without thermal noise, each initial configuration of disks can be uniquely mapped to a given jammed packing after specifying the packing-generation protocol~\cite{jamming:OHernPRE2003}. Therefore, in this study, we should consider similar sets of initial configurations for all four systems: attractive and repulsive bead-spring polymers and attractive and repulsive disks. To achieve the initial states, we generate liquid globule configurations of attractive bead-spring polymers. The initial disk configurations can be obtained from the liquid globules by replacing the bonded interactions with non-bonded interactions and the purely repulsive configurations can be obtained from the liquid globules by replacing the non-bonded attractive interactions with purely repulsive interactions. Packings at jamming onset for all four systems can then be generated through potential energy minimization using the appropriate potential energy functions described in Sec.~\ref{model}.

\subsubsection{Preparing initial configurations via polymer collapse}

\begin{figure}
\begin{center}
\includegraphics[width=0.875\columnwidth]{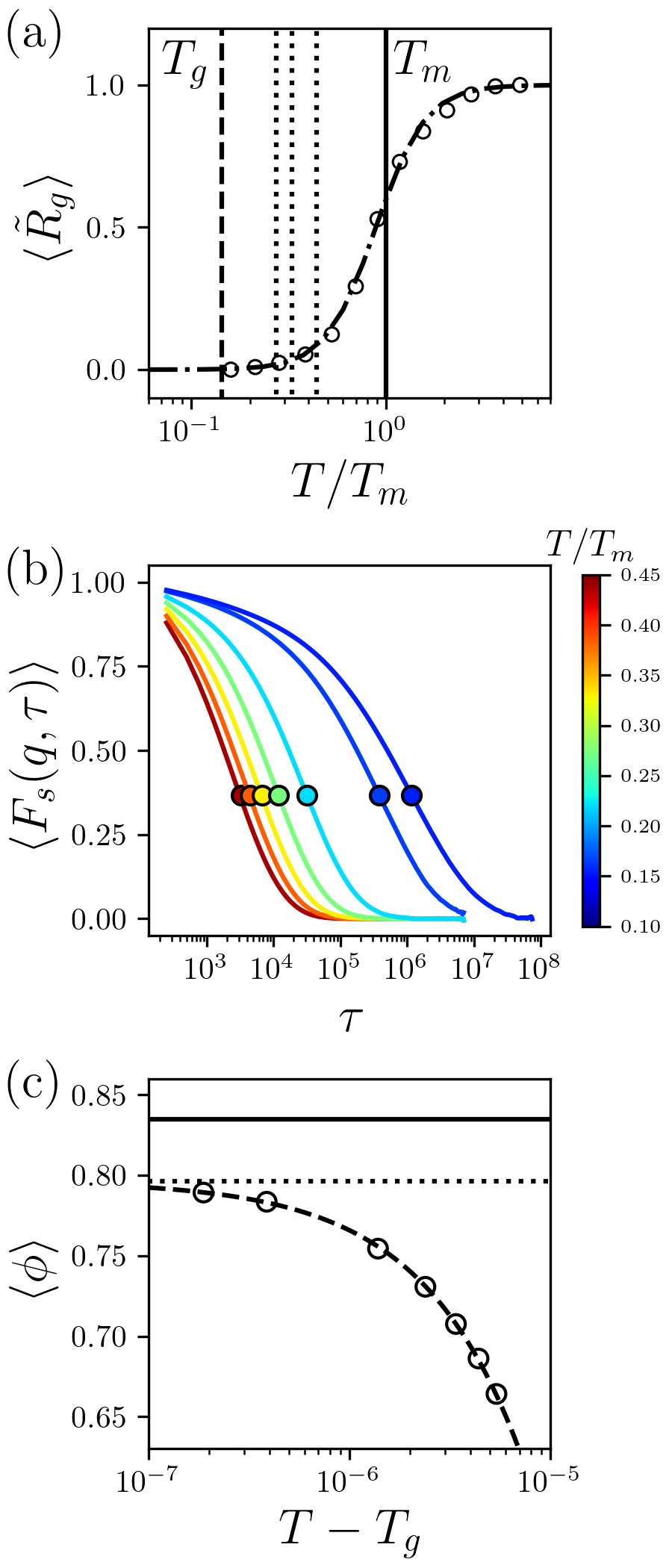}
\caption{(a) Normalized radius of gyration $\widetilde{R}_g$ plotted versus temperature $T$ normalized by the melting temperature $T_m$ (vertical solid black line). The dot-dashed line gives the fit of $\widetilde{R}_g$ to Eq.~\ref{eq:sigmoid}. (b) The self-part of the intermediate scattering function $F_s(q,t)$ at $q=2\pi/\sigma_{\rm max}$ averaged over all particles and time origins for several $T/T_m$. The filled circles indicate the structural relaxation times $\tau_r$ at which $F_s(q,\tau_r) = 1/e$. The colors from red to blue indicate high to low $T/T_m$. The vertical dashed line in (a) indicates $T_g$ below which $\tau_r \rightarrow \infty$. (c) The average core packing fraction $\langle \phi \rangle$ is plotted versus $T-T_g$. The dashed line gives $\langle \phi\rangle_g - \langle \phi \rangle \sim (T-T_g)^{\gamma}$, where $\langle \phi\rangle_g \approx 0.796$ (dotted line) and $\gamma \approx 0.9$. The horizontal solid line at $\langle \phi \rangle \approx 0.835$ indicates the average packing fraction at jamming onset for repulsive monomers under periodic boundary conditions. In all panels, the data are for attractive polymers and the angle brackets indicate averages over at least $10^2$ configurations generated via different initial conditions.}
\label{fig:phi_vs_TTg}
\end{center}
\end{figure}

To generate initial configurations, we simulate bead-spring polymers with non-bonded attractive interactions over a range of temperatures using a Langevin thermost. We integrate Newton's equations of motion for each monomer position ${\vec r}_j$ using a modified velocity-Verlet integration scheme with timestep $\Delta t = 0.01$~\cite{sim:Allen2017}. We characterize the temperature-dependent polymer configurations using the normalized radius of gyration:
\begin{equation}
\label{eq:isf}
    \widetilde{R}_g = \frac{R_g - R^{\rm min}_g}{R^{\rm max}_g-R^{\rm min}_g},
\end{equation}
where $R^{\rm max}_g$ and $R^{\rm min}_g$ are the maximum and minimum radii of gyration. As shown in Fig.~\ref{fig:phi_vs_TTg} (a) for $N=256$ and averaged over $100$ different initial conditions, 
polymers with attractive non-bonded interactions undergo two distinct transitions as they are cooled from high to low temperatures.
At high temperatures, the polymer samples an excluded-volume random walk. As the temperature is lowered, the attractive interactions overcome thermal fluctuations, and the polymer collapses into a condensed droplet, signaling the coil-to-globule transition. We can fit a sigmoidal curve to the normalized radius of gyration, 
\begin{equation}
\label{eq:sigmoid}
    \widetilde{R}_g(T) = \frac{1}{1+e^{\kappa (T-T_m)}},
\end{equation}
to identify the melting temperature $T_m$~\cite{polymercollapse:WilliamsAnnRev1981} at which $\widetilde{R}_g(T_m) = 1/2$ and where $\kappa$ gives the transition width. By cooling the polymer below $T_m$, we can induce a glass transition, where the structural relaxation time $\tau_r$ of the globule diverges. We determine $\tau_r$ by calculating the self-part of the intermediate scattering function,
\begin{equation}
    F_{s}(q,t) = \frac{1}{N} \left\langle \sum_{j=1}^{N} e^{ i \vec{q} \cdot (\vec{r}_j(t_0+t) - \vec{r}_j(t_0)) }\right\rangle,
\end{equation}
as a function of time $t$. The angle brackets indicate an average over time origins $t_0$ and directions of the wavenumber with magnitude $q=2\pi/\sigma_{\rm{max}}$. As shown in Fig.~\ref{fig:phi_vs_TTg} (b), at short times, $F_s(q,t) \sim 1$ since the monomer positions are similar to what they were at the time origin. $F_s(q,t)$ decays to zero when the configuration at time $t$ is uncorrelated with the initial configuration.  We define the structural relaxation time $\tau_r$ using $F_s(q,\tau_r) = 1/e$, which increases rapidly as the temperature decreases. We can estimate the glass transition temperature $T_g$ at which $\tau_r \rightarrow \infty$ using a power-law, $\tau_r \propto (T-T_g)^{-\lambda}$ (with $\lambda \approx 2$), or 
super-Arrhenius form, $\tau_r \propto e^{A/(T-T_g)}$ (with $A \approx 10$). Both forms give $T_g/T_m \approx 0.14$. The results in Fig.~\ref{fig:phi_vs_TTg} are shown for an interparticle potential with attractive range $\alpha = 1.5$ and depth $\beta = 10^{-5}$. We find qualitatively similar results for a range of $\alpha$ and $\beta$. Increasing $\beta$ shifts the melting curve to larger values of temperature, while increasing $\alpha$ broadens the coil-to-globule transition~\cite{sticky:PaulJCP2001}.

We first generate extended polymer configurations at high temperature $T \gg T_m$. We then slowly cool the polymers to temperatures $T_0$ below $T_m$, i.e. $T_0/T_m = 0.43$, $0.32$, and $0.27$, but above $T_g$, as shown in Fig.~\ref{fig:phi_vs_TTg} (a). We collect between $10^2$ and $10^3$ distinct sets of positions and velocities of the polymers at each $T_0$, with each set separated by $10\tau_r$. We consider $N=64$, $128$, $256$, $512$, and $1024$ to assess system-size effects. After generating the collapsed polymer configurations, we follow the protocols below to generate zero-temperature packings of polymers with non-bonded attractive interactions, disk packings with attractive interactions, packings of polymers with only non-bonded repulsive interactions, and disk packings with only repulsive interactions.

\subsubsection{Packing-generation protocol for attractive disks and polymers}

To generate static packings of attractive polymers, we cool liquid globules at $T_0$ to zero temperature using damped molecular dynamics (MD) simulations, where we solve Newton's equations of motion, 
\begin{equation}
\label{damped_MD}
m {\vec a}_j = -\partial V/\partial {\vec r}_j - b{\vec v}_j, 
\end{equation}
with dissipative forces proportional to the disk velocities ${\vec v}_j$, potential energy $V=V^b+V^{anb}$, disk mass $m$ and acceleration ${\vec a}_j$, and $j=1,\ldots,N$ labels the disks. For computational efficiency, each system is cooled using the reported damping parameter $b$ until the total force magnitude in the system reaches $F_{\rm tol}=\Sigma_{j=1}^N |\vec{F}_j| < 10^{-7}$, and then it is increased to $b=0.1$ in the overdamped limit. The simulations are terminated when $F_{\rm tol} < 10^{-15}$. 

The damped MD simulations can be performed on attractive disks (as well as attractive polymers) to investigate the effect of the polymer backbone on the zero-temperature packings. To generate static packings of attractive disks, we initialize the system with the positions and velocities of the collapsed globules at $T_0$ and then use damped MD simulations (Eq.~\ref{damped_MD}) to minimize the total potential energy, except now $V=V^{anb}$.

\subsubsection{Packing-generation protocol for purely repulsive disk and polymers}
\label{jamming}

For systems with attractive interactions, we employ open boundary conditions. Since static packings of purely repulsive particles possess non-zero pressures at jamming onset, they must be confined to form jammed packings, e.g. using periodic or fixed boundary conditions.  To generate jammed packings of purely repulsive particles in {\it open} boundary conditions, we include a linear spring potential that connects each particle to center of mass of the packing, which is the origin of the coordinate system, 
\begin{equation}
\label{eq:comp}
\frac{V^{c}(r_i)}{\epsilon} = \frac{k_{c}}{2\epsilon} r^2_i  \left( \sigma_i / \sigma_{\rm{max}} \right)^\nu,
\end{equation}
where $k_{c} \sigma_s^2 \ll \epsilon$ is the compressive energy scale. (See Appendix~\ref{sec:jamming_appendix} for a discussion of how the results depend on $k_{c}/\epsilon$.) To generate zero-temeprature packings of purely repulsive particles, we initialize the system with the positions and velocities from the collapsed globules at $T_0$. We then run damped MD simulations with $V=V^b + V^{rnb} + V^{c}$ for purely repulsive polymers or $V=V^{rnb} + V^{c}$ for purely repulsive monomers until force balance is achieved. The radial spring is then removed and the packings are again energy minimized until $F_{\rm tol} < 10^{-15}$.  For small damping coefficients, packings of repulsive disks with similar sizes segregate and crystallize. We thus include a factor of $\left( \sigma_i / \sigma_{\rm{max}} \right)^\nu$ with $\nu=2$ in Eq.~\ref{eq:comp} to prevent size segregation. (See Appendix~\ref{sec:jamming_appendix}.) 

To calculate the structural and mechanical properties of the packings as a function of the packing fraction above jamming onset, we add a repulsive circular boundary with radius $R$ via the repulsive linear spring potential,
\begin{equation}
\label{eq:sbc}
    \frac{V^{w}(r_i)}{\epsilon} = \frac{1}{2} \left( 1-\frac{R-r_{i}}{\sigma_{i}} \right)^2 \Theta \left(1-\frac{R-r_{i}}{\sigma_{i}} \right).
\end{equation}
$R$ is initialized so that there are no disk-wall contacts. The system is successively compressed by scaling the wall and particle positions such that $r_{i}' = r_i (1 - 2\Delta \phi/\phi)$ with each compression step $\Delta \phi =10^{-3}$ followed by energy minimization using damped MD simulations with $b=0.1$. The system is compressed until it reaches a target total potential energy per particle $V_0 < V/N < 2V_0$. If the system is compressed above $V/N > 2V_0$, the previous particle positions and boundary radius are re-initialized, the system is compressed by $\Delta \phi / 2$, and energy-minimized. The static packings were prepared over a wide range of potential energies per particle, $10^{-13} \lesssim V_0 \lesssim 10^{-2}$.

\subsection{Core packing fraction}

To analyze the structural properties of the interiors of static packings, their surfaces must first be identified. To do this, we adapt and apply an algorithm first proposed for finding the surfaces of proteins in solvent from Lee and Richards~\cite{rsasa:LeeJMB1971}. We first place a probe disk of diameter $\sigma_p$ on the surface of the disk or polymer packing. It is then rolled over the surface of the packing until it returns to its initial location. In this study, we consider any disk touched by the probe as a `surface' disk. The size of the probe disk affects which disks are considered as surface disks. We set $\sigma_p/\sigma_s = 0.1$, which is similar to the ratio of the diameter of a water molecule to the diameter of alanine. The variation of the average core packing fraction in static packings with $\sigma_p/\sigma_s$ is investigated in Appendix~\ref{sec:rsasa_appendix}.

After identifying the surface disks of a given configuration, a radical Voronoi tessellation is performed on the disk centers within a square box with an edge length exceeding the largest extent of each packing~\cite{voro++:Rycroft2009}. The core packing fraction for a particular configuration is defined as 
\begin{equation}
    \phi =  \frac{\sum_{i=1}^{N_c}\pi r_i^2}{\sum_{i=1}^{N_c}a_i},
\end{equation}
where $N_c$ is the number of core disks and $a_i$ the area of the Voronoi polygon surrounding the $i$th core disk. Due to the small probe radius, all of the core disks have closed Voronoi cells and so their areas do not depend on the enclosing box size.

\section{Results}
\label{sec:results}

\begin{figure}
\begin{center}
\includegraphics[width=0.875\columnwidth]{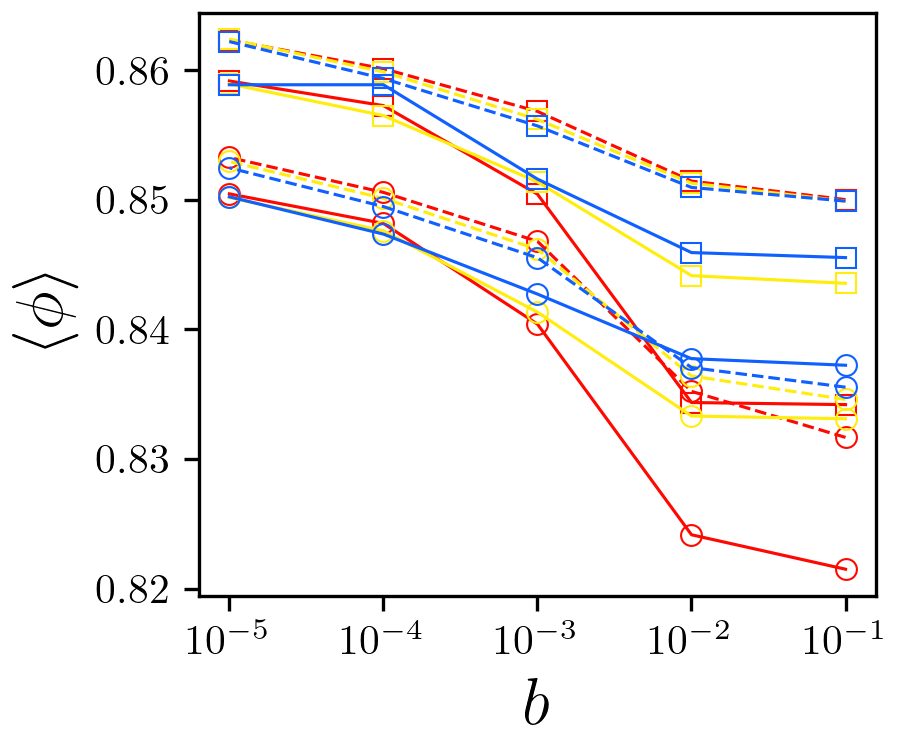}
\caption{The average core packing fraction $\langle \phi \rangle$ from damped MD simulations plotted versus the damping parameter $b$ for attractive disk-shaped bead-spring polymers (circles with solid lines), attractive disks (squares with solid lines), repulsive disk-shaped bead-spring polymers (circles with dashed lines), and repulsive disks (squares with dashed lines), prepared from initial temperatures $T_0/T_m = 0.43$ (red), $0.32$ (yellow), and $0.27$ (blue) for $N=512$.}
\label{fig:packings}
\end{center}
\end{figure}

In this section, we describe the structural and mechanical properties of static packings of disks and disk-shaped bead-spring polymers with purely repulsive, as well as attractive interactions. In Sec.~\ref{sec:results_Tg}, we first show that when attractive disk-shaped bead-spring polymers are cooled toward the glass transition temperature $T_g$, the average packing fraction of the interior (or core region) is well-below values given for random close packing for disordered packings of repulsive disks. Therefore, in Sec.~\ref{sec:results_core} we study the core packing fraction of attractive polymers as they are cooled from $T_0 > T_g$ to zero temperature using damped MD simulations. We find that attractive disk-shaped bead-spring polymers, as well as attractive disks, when cooled to zero temperature, possess similar core packing fractions as found for static packings of repulsive disks and disk-shaped bead-spring polymers over a wide range of initial temperatures $T_0$, damping parameters $b$, and system sizes $N$. In Sec.~\ref{sec:results_vdos}, we show that attractive disks and disk-shaped bead-spring polymers quenched to zero temperature possess an excess number of low-frequency modes in the density of vibrational states (similar to jammed packings of repulsive disks). We further show that slowly increasing the depth $\beta$ of the attractive interparticle potential causes the attractive packings to lose low-frequency modes in a way that is similar to compression of repulsive disk packings above jamming onset. In Sec.~\ref{sec:results_hypo}, we find that, contrary to previous studies, static packings of repulsive disk-shaped bead-spring polymers are hypostatic at jamming onset, but the number of missing contacts relative to the isostatic number matches the number of quartic modes that arise from the polymer backbone constraints. When we account for the quartic modes, the excess number of contacts above isostaticity (for packings of repulsive polymers) scales as $\Delta N \sim \left( V_rN^3 \right)^\alpha$, where $V_r$ is the total repulsive potential energy of the packing, $\alpha=1/2$ at small $\Delta N$, and the exponent crosses over to $\alpha=1/4$ in the large-$\Delta N$ limit. Finally, in Sec.~\ref{sec:results_iso} we show that zero-temperature attractive disks and disk-shaped bead-spring polymers are also effectively isostatic if contacts are defined as $r_{ij} < r_\beta$ and they obey the same scaling of the excess number of contacts with the repulsive energy, $\Delta N \sim \left( V_r N^3 \right)^{\alpha}$, as found for static packings of repulsive disks and disk-shaped bead-spring polymers.

\subsection{Core packing fraction for collapsed polymers near $T_g$ is well below random close packing for repulsive disks}
\label{sec:results_Tg}

What is the core packing fraction of an attractive disk-shaped bead-spring polymer as it is cooled toward the glass transition temperature $T_g$? In Fig.~\ref{fig:phi_vs_TTg} (c), we plot the average core packing fraction $\langle \phi \rangle$ versus $T-T_g$ for $N=256$ averaged over $100$ polymers with different initial conditions. The core packing fraction increases with decreasing temperature, $\langle \phi\rangle_g - \langle \phi \rangle \sim (T-T_g)^{\gamma}$, approaching the plateau value of $\langle \phi \rangle_g \approx 0.796$ as $T \rightarrow T_g$ (with $\gamma \approx 0.9$). $\langle \phi \rangle_g$ is similar to values reported for the packing fraction near the glass transition in experimental, computational, and theoretical studies of hard spheres~\cite{glass:VivekPNAS2017,glass:LiaoPNAS2023}. In contrast, static packings of $N=256$ purely repulsive polydisperse disks, without a polymer backbone, possess a much larger packing fraction, $\langle \phi \rangle \approx 0.835$, at jamming onset~\cite{jamming:OHernPRE2003}. The core packing fraction for collapsed attractive polymers near $T_g$ is far below that for static packings of purely repulsive disks at jamming onset. This result indicates that for the core packing fraction of collapsed attractive polymers to reach those of jammed disconnected, repulsive disks, they must be cooled to temperatures much below the glass transition temperature.

\subsection{Core packing fraction for collapsed polymers with $T \ll T_g$ matches that for jammed repulsive disk packings}
\label{sec:results_core}

To study the core packing fraction of collapsed, attractive polymers below the glass transition temperature $T_g$, we performed damped MD simulations to take attractive polymers with initial temperatures $T_m > T_0 > T_g$ to zero temperature using a wide range of damping parameters. In Fig.~\ref{fig:packings}, we show that the core packing fraction of collapsed, attractive polymers increases with decreasing damping parameter from roughly $0.83$-$0.84$ to $0.85$ (circles with solid lines) for $N=512$. For large damping parameters, larger initial temperatures $T_0$ give rise to the lowest values of the core packing fraction. However, for low damping parameters, the results for the core packing fraction of collapsed, attractive polymers are the same for all $T_0$. To study the effects of the polymer backbone constraint on the core packing fraction, we repeat these simulations for disconnected, attractive disks (squares with solid lines). The dependence of $\langle \phi \rangle$ on the damping parameter $b$ and initial temperature $T_0$ is similar to that for collapsed, attractive polymers, however, the packing fraction is shifted to larger values by $\approx 0.01$ for all $b$ and $T_0$.

\begin{figure}
\begin{center}
\includegraphics[width=0.875\columnwidth]{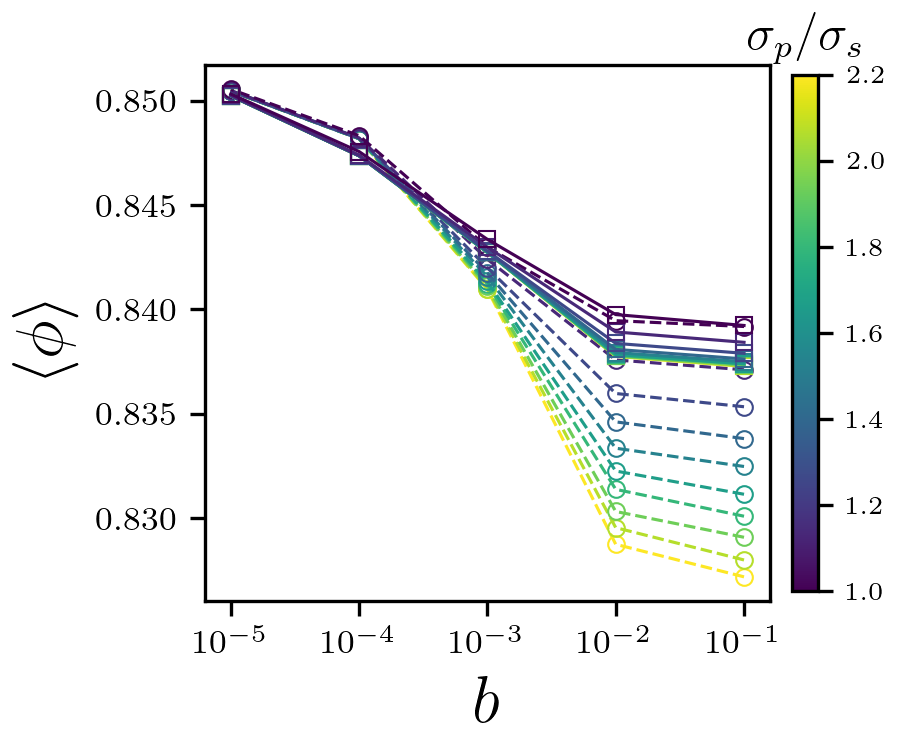}
\caption{The average core packing fraction $\langle \phi \rangle$ from damped MD simulations of attractive polymers initialized at $T_0/T_m = 0.43$ (circles with dashed lines) and $0.27$ (squares with dashed lines) plotted versus the damping parameter $b$ when void regions are identified using probe diameters, $2.2 \lesssim \sigma_p/\sigma_s < 1$ (where purple to yellow indicates increasing size), for $N=512$. Core disks adjacent to void regions are not included in the calculation of $\langle \phi\rangle$.}
\label{fig:holes}
\end{center}
\end{figure}

To compare the core packing fraction of collapsed, attractive polymers to the packing fraction of jammed repulsive systems, we developed a novel compression protocol to generate jammed repulsive systems in open boundary conditions. (See Sec.~\ref{jamming}.) We start with the same attractive polymer configurations prepared at $T_0$ for both polymers and disconnected disks. We then replace the non-bonded attractive interactions ($V^{anb}$) with non-bonded repulsive interactions ($V^{rnb}$) and compress the system isotropically by attaching each disk to a radial linear spring anchored to the origin. In Fig.~\ref{fig:packings}, we show the core packing fraction for jammed packings of repulsive disk-shaped bead-spring polymers (circles with dashed lines) and repulsive disks (squares with dashed lines). For these purely repulsive systems, the core packing fraction does not depend strongly on $T_0$. Further, for small $T_0$, the collapsed, attractive polymers and jammed repulsive polymers possess similar core packing fractions for all damping parameters $b$. In addition, there is qualitative agreement for the core packing fraction of packings of disconnected attractive and repulsive disks for all $b$. These results emphasize that the attractive interactions do not strongly influence the core packing fraction, i.e. structures that collapse due to attractive interactions are similar to those that form due to mechanical compression with weak thermal 
fluctuations. 

As discussed above, the core packing fraction for collapsed, attractive polymers is the lowest for large damping parameters $b$ and high initial temperatures $T_0$. We find that these collapsed structures possess large void regions surround by regions that are densely packed. To identify the void regions, we test each interior disk to determine whether a probe disk of diameter $\sigma_p$ can be placed at its edge without causing any overlaps. If the probe can be placed without causing overlaps, we remove that disk from the list of core disks. In Fig.~\ref{fig:holes}, we show that when we remove core disks that are near void regions (by choosing $\sigma_p/\sigma_s=1$), the core packing fraction $\langle \phi \rangle$ is no longer strongly dependent on $T_0$ for large damping parameters. Since the collapsed structures in the low-damping limit do not possess void regions, $\langle \phi \rangle$ does not depend on $T_0$ or $\sigma_p$ for small $b$. Thus, aside from void regions, the initial temperature has only a minor effect on the packing fraction of dense core regions of collapsed, attractive polymers. 

\begin{figure}
\begin{center}
\includegraphics[width=0.875\columnwidth]{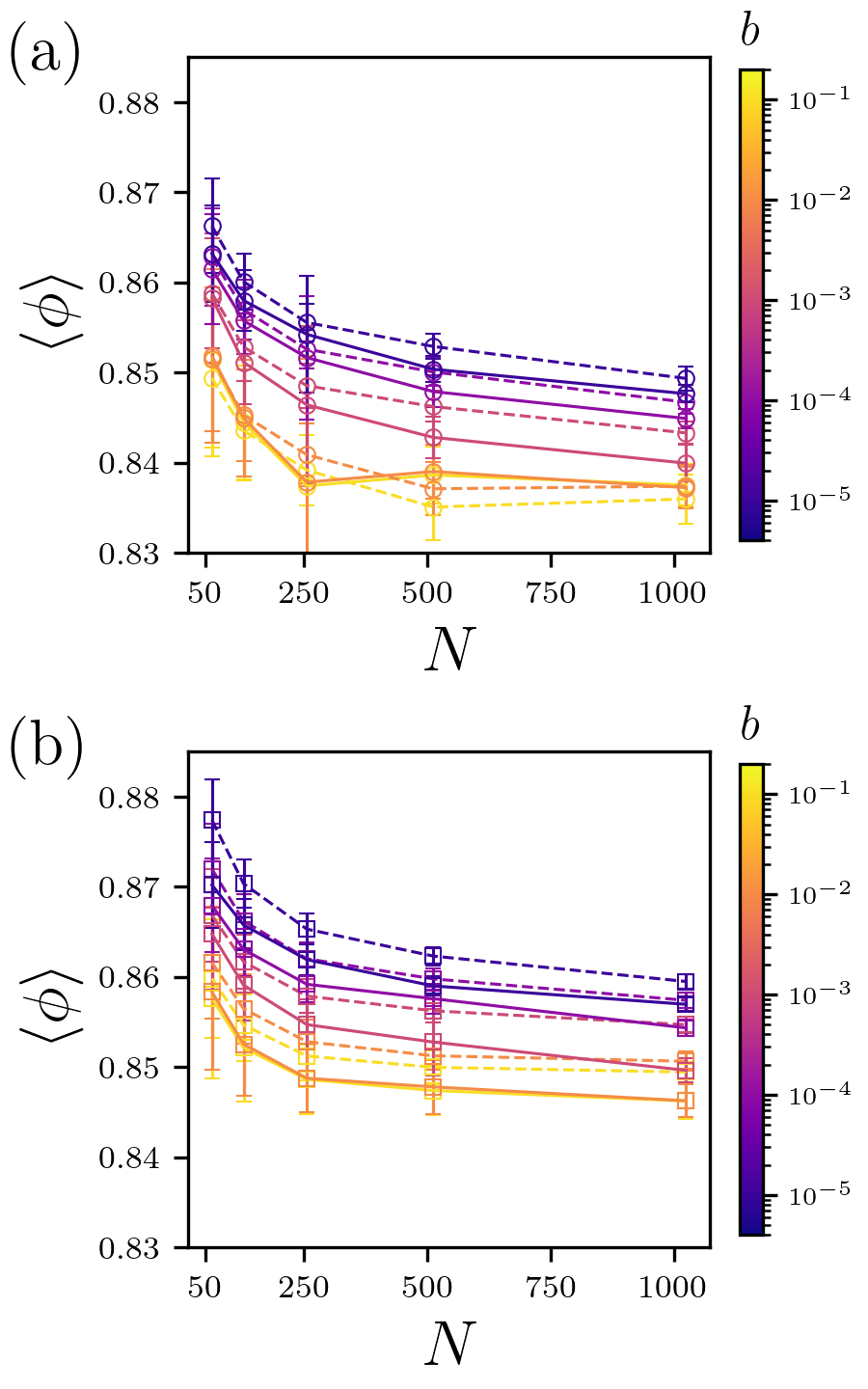}
\caption{The core packing fraction $\langle \phi \rangle$ from damped MD simulations averaged over all initial temperatures $T_0$ and plotted versus the system size $N$ and damping parameter $b$ (increasing from purple to yellow). We show results for (a) collapsed, attractive polymers (circles with solid lines) and jammed repulsive polymers (circles with dashed lines) and (b) attractive disks (squares with solid lines) and jammed repulsive disks (squares with dashed lines). Void regions are identified using probe size $\sigma_p=1$ and core disks adjacent to void regions are not included in the calculation of $\langle \phi\rangle$. }
\label{fig:packing_no_holes}
\end{center}
\end{figure}

In Fig.~\ref{fig:packing_no_holes}, we present the results for the core packing fraction (averaged over all $T_0$ and excluding void regions) plotted versus the system size $N$ and damping parameter $b$ for (a) disk-shaped bead-spring polymers and (b) disconnected disks. In general, when we do not consider void regions, the core packing fraction for collapsed, attractive polymers matches that for jammed, repulsive polymers and the core packing fraction for packings of attractive disks matches that for jammed repulsive disks for all $b$ and $N$. These results suggest that the structural properties of systems with attractive interactions that are cooled to zero temperature are similar to those for repulsive systems that are compressed to jamming onset. In addition, we find that the average core packing fraction \emph{decreases} with increasing system size $N$, whereas packing-generation protocols that start from low-density configurations yield $\langle \phi \rangle$ that typically \emph{increase} with $N$~\cite{jamming:OHernPRE2003}.  For polymers, $\langle \phi \rangle$ varies between $0.84$-$0.85$ in the large-$N$ limit. For disks, $\langle \phi \rangle \approx 0.85$-$0.86$ for large $N$. 

\begin{figure}
\begin{center}
\includegraphics[width=0.875\columnwidth]{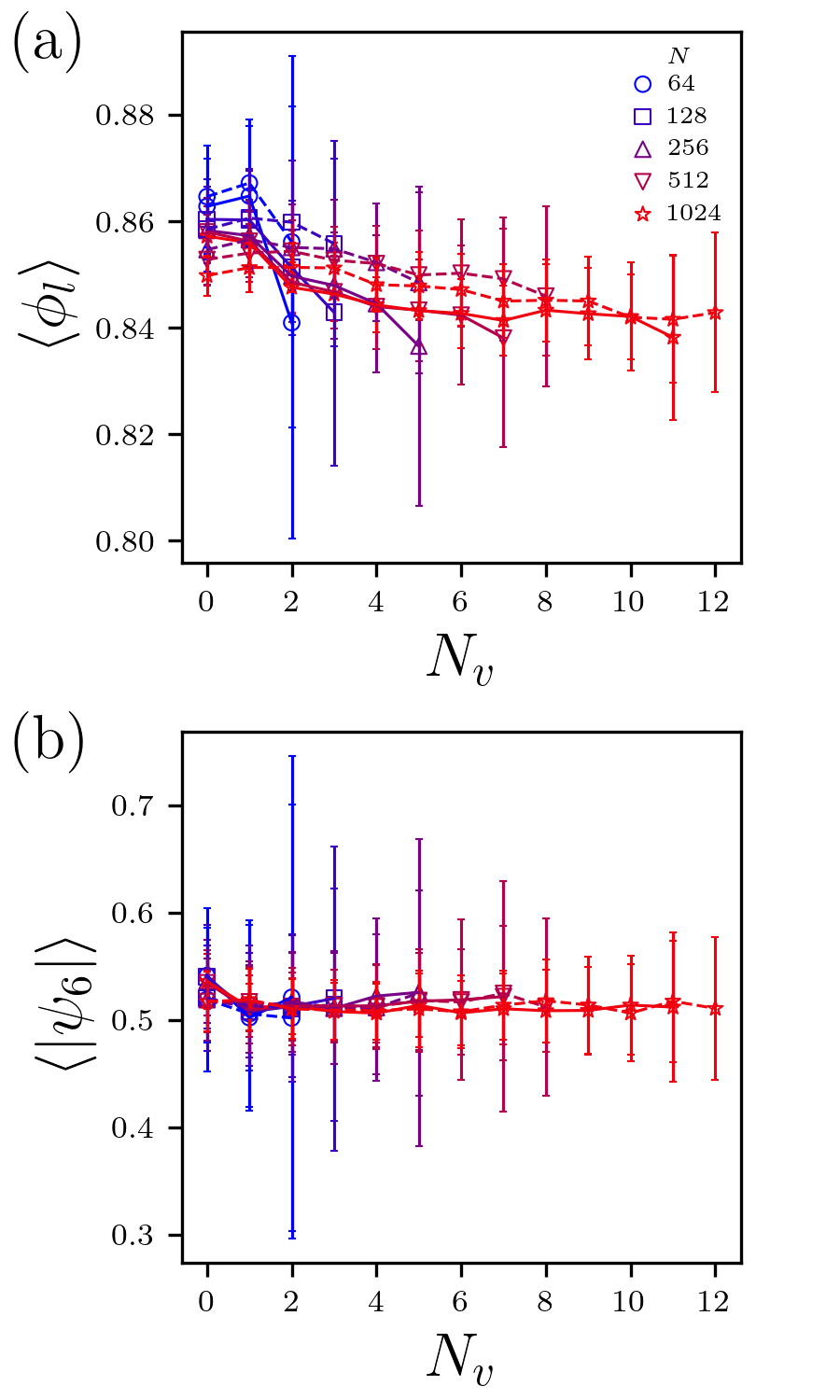}
\caption{(a) The local packing fraction $\langle \phi_l \rangle$ and (b) hexatic order parameter $\langle \vert \psi_6 \vert \rangle$ for each disk plotted versus the number of Voronoi cells $N_{\nu}$ between each disk and the closest surface disk for collapsed, attractive polymers (solid lines) and jammed, repulsive polymers (dashed lines) for several system sizes, $N=64$ (circles), $128$ (squares), $256$ (upward triangles), $512$ (downward triangles), and $1024$ (stars).}
\label{fig:packing_vs_nv}
\end{center}
\end{figure}

To better understand the system-size dependence of $\langle \phi \rangle$, we also calculate the local core packing fraction $\phi_l$ as a function of the distance to the surface of the packing. For small packings, a relatively large fraction of the disks are located near the curved boundaries. As $N$ increases, a larger number of disks are considered bulk, far from the curved boundaries. In Fig.~~\ref{fig:packing_vs_nv} (a), we plot the local core packing fraction $\phi_l$ versus the number of Voronoi cells $N_{\nu}$ between a given disk and the closest surface disk for collapsed, attractive polymers and jammed, repulsive polymers. ($N_{\nu}=0$ indicates that a core disk is adjacent to a surface disk.) We find that the core packing fraction for both attractive and repulsive polymers is largest for small systems and near surface disks. As $N_{\nu}$ increases, $\langle \phi_l \rangle$ decreases and converges in the large-system limit. In addition, $\langle \phi_l\rangle$ is more uniform for jammed, repulsive polymer packings. 

We also calculated the local hexatic order parameter associated with each core disk,
\begin{equation}
\label{eq:psi_6}
    |\psi_6| = \frac{1}{n_k} \left|\sum_{j=1}^{n_k} e^{6i\theta_{jk}} \right|,
\end{equation}
where $\theta_{jk}$ is the angle between a central core disk $k$ and its Voronoi neighbors $j=1$,$\ldots$,$n_k$,  
to determine whether increases in the core packing fraction are correlated with increases in positional order. In Fig.~\ref{fig:packing_vs_nv} (b), we show that $\langle \vert \psi_6 \vert \rangle \sim 0.5$ is independent of $N_{\nu}$ and comparable to values for amorphous jammed disk packings~\cite{jamming:SchreckPRE2011}.

\subsection{Low-frequency contribution to the density of vibrational modes}
\label{sec:results_vdos}

\begin{figure*}
\begin{center}
\includegraphics[width=0.95\textwidth]{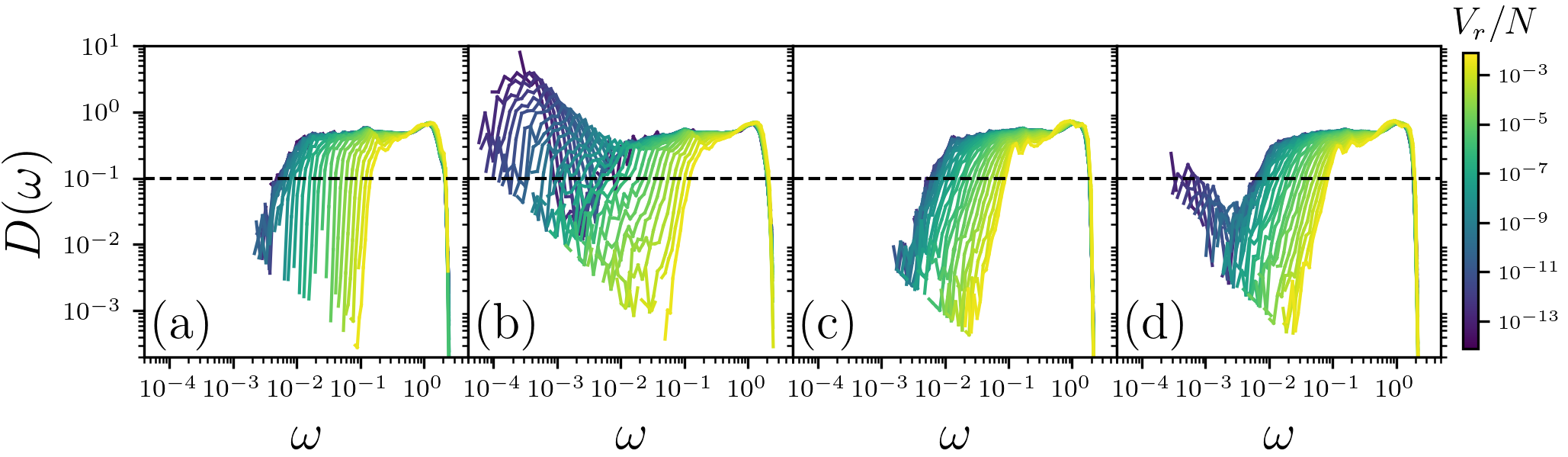}
\caption{The vibrational density of states $D(\omega)$ for (a) jammed repulsive disks, (b) jammed repulsive polymers, (c) attractive disks, and (d) attractive polymers, colored by $V_r/N$ (increasing from purple to yellow) for $N=128$. The black dashed line defines the characteristic frequency $\omega^{\ast}$, where $D(\omega^{\ast})=10^{-1}$. Note the large low-frequency peak for packings of repulsive and attractive  polymers in (b) and (d), which arise due to quartic modes. Quartic modes are removed from $D(\omega)$ when calculating $\omega^{\ast}$. (See Sec.~\ref{sec:results_hypo}.)}
\label{fig:vdos}
\end{center}
\end{figure*}

Above, we showed that the core packing fractions for collapsed, attractive polymers and packings of attractive disks are similar to those of jammed repulsive polymers and repulsive disks. Do these disparate systems also share the other structural and mechanical properties of jammed packings of repulsive disks? We first consider the vibrational density of states $D(\omega)$, which is obtained by calculating the dynamical matrix, 
\begin{equation}
\label{eq:DM}
    M_{kl} = \frac{\partial^2 V}{\partial \vec{r}_k \partial \vec{r}_l},
\end{equation}
where $k$ and $l$ label the $2N$ coordinates of the disks. The eigenvectors $\vec{\xi}^i_k = \{e^i_{1x}, e^i_{1y}, \ldots,e^i_{Nx}, e^i_{Ny} \}$ represent an orthogonal set of $2N$ normal modes whose eigenvalues $e^i$ correspond to the normal mode frequencies $\omega^i=\sqrt{e^i}$. $D(\omega)$ does not depend strongly on the initial temperature $T_0$ or the damping paramter $b$ used to generate the packings, and we focus on packings prepared using $T_0/T_m = 0.27$ and $b=10^{-5}$. To generate mechanically stable repulsive packings, we jammed the repulsive disks and polymers under circular boundary conditions. Specifically, we initialize the repulsive packings analyzed in Sec.~\ref{sec:results_core} and then apply sequential affine compressions of $\Delta \phi = 10^{-3}$ followed by overdamped energy minimization until reaching a target potential energy $V_r/N=10^{-14}$, where $V_r=V^{rnb}+V^{w}$ for repulsive disks and $V_r=V^{rnb}+V^{b}+V^{w}$ for repulsive polymers. Additionally, underconstrained disks associated with zero-modes are removed---rattlers in the case of repulsive disks and flippers in the case of repulsive polymers. (See Sec.~\ref{sec:results_hypo} for further details.) In Fig.~\ref{fig:vdos} (a) and (b), we show the density of vibrational states $D(\omega)$ for packings of repulsive disks and packings of repulsive polymers, respectively. As expected, $D(\omega)$ for jammed packings of repulsive disks possess an anomalous plateau at low frequencies rather than Debye behavior~\cite{jamming:OHernPRE2003}. Similarly, packings of repulsive polymers also display a low-frequency plateau with $10^{-2} < \omega < 10^{-1}$ in Fig.~\ref{fig:vdos} (b). However, there are further excess vibrational modes in packings of repulsive polymers for $\omega < 10^{-2}$, which indicate the presence of quartic modes that are discussed below in Sec.~\ref{sec:results_hypo}.

When the attractive interactions are weak, i.e. $\beta=10^{-5}$ as discussed in Sec.~\ref{sec:results_core}, attractive disk and polymer packings possess only small disk overlaps, $V_r / N \lesssim 10^{-14}$, where $V_r = V^{rnb}$ for attractive disks and $V_r = V^{nrb}+V^{b}$ for attractive polymers. We find that $D(\omega)$ for attractive disk and attractive polymer packings with $V_r / N \lesssim 10^{-14}$ possess no non-trivial zero modes and a broad low-frequency plateau, similar to that obtained for jammed, repulsive disk packings prepared with comparable values of $V_r$ as shown in Fig.~\ref{fig:vdos} (c) and (d). The small peak at the lowest frequencies in packings of attractive polymers indicates the presence of quartic modes. 

\begin{figure*}
\begin{center}
\includegraphics[width=0.95\textwidth]{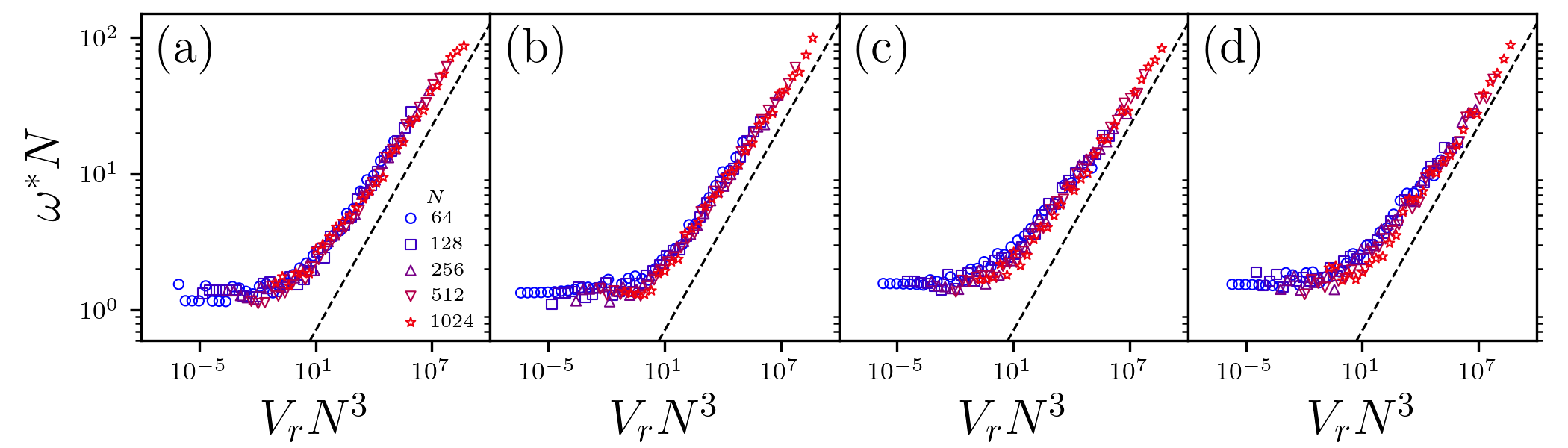}
\caption{Characteristic plateau frequency of the vibrational density of states $\omega^{\ast} N$ versus potential energy $V_r N^3$ for packings of (a) repulsive disks, (b) repulsive polymers, (c) attractive disks, and (d) attractive polymers as a function of system size, $N=64$ (circles), $128$ (squares), $256$ (upward triangles), $512$ (downward triangles), and $1024$ (stars) colored from blue to red with increasing system size. The dashed line has a slope of $0.25$.}
\label{fig:U_vs_omega}
\end{center}
\end{figure*}

When we compress repulsive disk and polymer packings above jamming onset by increasing $\phi$ and thus $V_r$ (from purple to yellow), the plateau in $D(\omega)$ at low frequencies decreases, as shown in Fig~\ref{fig:vdos} (a) and (b) ~\cite{vdos:SilbertPRL2005,vdos:WyartPRE2005}. Effective compression of attractive packings can be obtained by increasing the attractive depth $\beta$. In Fig.~\ref{fig:vdos} (c) and (d), we vary the attractive depth by successively multiplying $\beta$ by a factor of $r \sim 1.12$ in the range $10^{-8} < \beta < 10^{-1}$ followed by overdamped energy minimization after each change in $\beta$.  Increasing $\beta$ gives rise to concomitant increases in $V_r$ and a loss of the low-frequency plateau. 

We quantify the anomalous low-frequency plateau in $D(\omega)$ by identifying a characteristic frequency $\omega^{\ast}$ at which $D(\omega^{\ast})$ falls below a small threshold. Here, we use $D(\omega^{\ast})=10^{-1}$, but the results are similar over a range of thresholds. In Fig.~\ref{fig:U_vs_omega} (a), we show $\omega^{\ast}$ as a function of $V_r$ for packings of repulsive disks compressed under circular boundary conditions for several system sizes $N=64$, $128$, $256$, $512$, and $1024$. Previous work has shown that under periodic boundary conditions the characteristic plateau frequency scales as $\omega^{\ast} N \sim \left( PN^2 \right)^{1/2}$ at high pressures $P$~\cite{vdos:SilbertPRL2005,vdos:WyartPRE2005,contacts:GoodrichPRE2014}. Attractive packings with no boundaries are at zero pressure, and thus we plot their low frequency response against $V_r$ instead of $P$. Potential energy $V$ and pressure $P$ in repulsive systems have a known scaling relation of $P \sim \left( V/N \right)^{1/2}$~\cite{jamming:OHernPRE2003}. Combining these two scaling relations gives $\omega^{\ast} N \sim \left( VN^{3} \right)^{1/4}$, which is plotted as black dashed line in Fig.~\ref{fig:U_vs_omega} (a)~\cite{jamming:WuPRE2017}. Additionally, we show in Fig.~\ref{fig:U_vs_omega} (b) that compressing repulsive polymer packings above jamming onset gives nearly identical results for $\omega^{\ast}N$ versus $V_r N^3$ as found for repulsive disk packings, when quartic modes are removed. This result indicates at least in the harmonic approximation double-sided polymer bonds do not strongly affect the low-frequency mechanical response.

Does the power-law scaling of $\omega^{\ast}$ versus $V_r$ still hold for attractive packings as we increase $\beta$ and thus $V_r$? In Fig.~\ref{fig:U_vs_omega} (c) and (d), we show that increasing the attraction depth is similar to overcompression of a repulsive disk packing, i.e. both lead to a decrease in the low-frequency plateau in $D(\omega)$ and give rise to $\omega^{\ast} N \sim (V_r N^3)^{1/4}$ for the finite-size scaling of the plateau frequency. In Fig.~\ref{fig:U_vs_omega}, we achieved an effective compression of attractive packings by increasing the attractive depth $\beta$, while fixing the attractive interaction range at $\alpha = 1.5$. In Sec.~\ref{sec:results_iso} we address varying $\alpha$ as well as $\beta$ and find similar results.

\subsection{Repulsive polymer packings are hypostatic, but effectively isostatic}
\label{sec:results_hypo}

\begin{figure*}
\begin{center}
\includegraphics[width=0.95\textwidth]{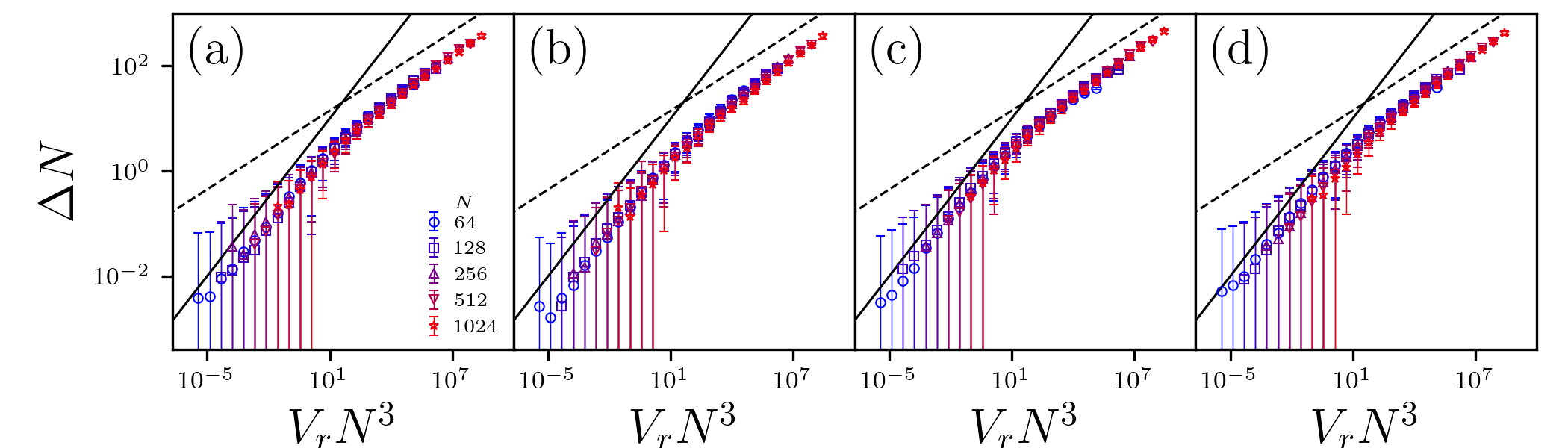}
\caption{Excess contact number above isostaticity $\Delta N=N_c' - N_c^{\rm iso}$ versus potential energy $V_r N^{3}$ for packings of (a) repulsive disks $(N_c'= N_c + N_w)$, (b) repulsive polymers $(N_c'= N_c + N_w + N_b + N_q)$, (c) attractive disks $(N_c'= N_c(r_{ij}<r_{\beta}) + N_b)$ and (d) attractive polymers $(N_c'= N_c(r_{ij}<r_{\beta}) + N_b + N_q)$ as a function of system size, $N=64$ (circles), $128$ (squares), $256$ (upward triangles), $512$ (downward triangles), and $1024$ (stars) colored from blue to red with increasing system size. $N_c$ is the number of interparticle contacts, $N_w$ is the number of particle-wall contacts, $N_b$ is the number of polymer bonds, and $N_q$ is the number of quartic modes. The solid line indicates slope $0.5$ and the dashed line indicates slope $0.25$. Error bars indicate one standard deviation in $\Delta N$.}
\label{fig:U_vs_delta_N}
\end{center}
\end{figure*}

Jammed packings of repulsive disks are known to be isostatic, i.e. the onset of rigidity occurs when the number of constraints (arising from interparticle and particle-wall contacts) equals the number of degrees of freedom. For isostatic packings, the number of contacts at jamming onset satisfies: $N^{\rm iso}_c=2(N-N_r)+f(d)+1$, where $N_r$ is the number of underconstrained rattler particles, $f(d)$ indicates the number of unconstrained degrees of freedom from the boundary conditions (e.g. $f(d)=1$ for circular fixed boundaries in $d=2$), and the $+1$ corresponds to the particle size degree of freedom~\cite{jamming:MaksePRL2000,jamming:OHernPRE2003}. Rattler particles for packings of repulsive disks correspond to particles with fewer than three contacts or particles where all contacts occur on a semicircle. Rattler particles are identified and removed iteratively. Previous studies have shown that compressing jammed packings gives rise to an increase in interparticle contacts, which in turn increases the characteristic plateau frequency $\omega^{\ast}$. In Fig~\ref{fig:U_vs_delta_N} (a), we plot $\Delta N = N_c + N_w - N_c^{\rm iso}$ versus $V_r N^3$, where $N_c$ is the number of interparticle 
contacts and $N_w$ is the number of particle-wall contacts. We show that $\Delta N$ obeys power-law scaling with $V_r/N$: $\Delta N \sim (V_r N^3)^{\zeta}$, where $\zeta = 0.5$ for $V_r N^4 \lesssim 1$ and $\zeta = 0.25$ for $V_r N^3 \gtrsim 1$.  These results match those for the finite-size scaling of the pressure dependence of $\Delta N$ and shear modulus $G$ for jammed packings of repulsive disks and spheres~\cite{contacts:GoodrichPRE2014,jamming:WangPRE2021}, i.e. $\Delta N \sim G \sim (pN^2)^{\lambda}$, where $\lambda = 1$ for $pN^2 \lesssim 1$ and $\lambda = 0.5$ for $pN^2 \gtrsim 1$.

\begin{figure}
\begin{center}
\includegraphics[width=0.875\columnwidth]{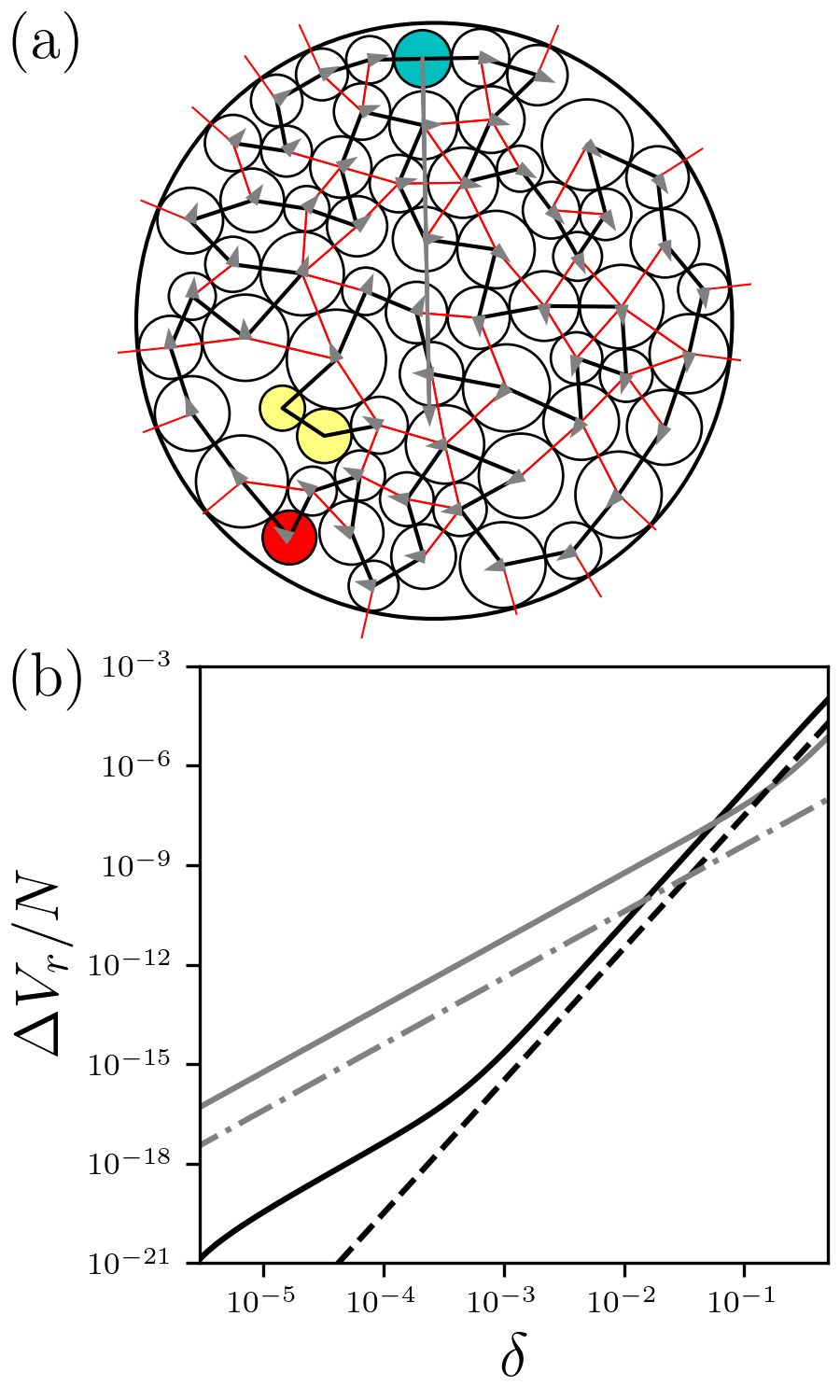}
\caption{(a) Jammed repulsive polymer packing showing the quartic mode in (b) with grey arrows for $N=64$. Red lines indicate interparticle and particle-wall contacts. Black lines indicate the polymer backbone. The large black circle that encloses the polymer indicates the circular wall. Non-flipper disks are colored white. The pair of yellow disks are underconstrained flippers. The cyan disk has no non-bonded contacts and participates most directly in the quartic mode. The red disk also has no non-bonded contacts, but does not lead to a quartic mode. (b) Change in potential energy $\Delta V_r/N$ following a perturbation with amplitude $\delta$ applied along an eigenvector of the dynamical matrix for a jammed repulsive polymer packing corresponding to a quadratic (grey solid line) and quartic mode (black solid line). Grey dot-dashed and black dashed lines indicate slopes of $2$ and $4$.}
\label{fig:quaritc}
\end{center}
\end{figure}

Previous studies have suggested that jammed packings of repulsive polymers are isostatic~\cite{packing:KarayiannisJCP2009,packing:SoikPRE2019}. However, one must carefully identify ``flipper'' particles that have too few contacts to be fully constrained, as well as quartic modes. We find that jammed packings of repulsive polymers are in fact {\it hypostatic}, but are effectively isostatic when accounting for flippers and quartic modes. Previous work identified flipper particles as those with no non-bonded interactions~\cite{packing:KarayiannisPRL2008,packing:SoikPRE2019}. Here, we use (non-rotational) zero modes of the dynamical matrix ${\vec \xi}^i$ to identify underconstrained flipper particles in repulsive polymer packings. We successively remove the largest contribution $\{ e^i_{jx},e^i_{jy} \}$ to ${\vec \xi}^i$ until it is no longer a zero mode. Each particle $j$ with the largest contribution to the zero mode is identified as a flipper particle. In Fig.~\ref{fig:quaritc} (a), the yellow-shaded particles are flippers since they only have bonded contacts, one of their neighbors only has bonded contacts, and they can collectively rotate without changing the length of the bonds and without making additional contacts. The red and cyan particles have no non-bonded contacts, but their bonded neighbors have at least one non-bonded contact, and so they are not flipper particles.

The grey arrows in Fig.~\ref{fig:quaritc} (a) indicate a quartic mode in a repulsive polymer packing. The cyan particle has the largest contribution to the quartic mode and its motion is perpendicular to the approximately $180 \degree$ bond angle. When we perturb a packing by an amplitude $\delta$ along a typical eigenvector ${\vec \xi}^i$ of the dynamical matrix, the change in potential energy $\Delta V_r \sim \delta^2$ scales quadratically with the amplitude as shown in Fig.~\ref{fig:quaritc} (b). However, hypostatic packings contain quartic modes, such that the change in energy $\Delta V_r$ for perturbations with amplitude $\delta$ along a quartic mode scale as $\Delta V_r \sim \delta^4$~\cite{stiffness:SchreckPRE2012}. In Fig.~\ref{fig:quaritc} (b), we show the quartic scaling for $\delta \gtrsim \delta_q$, where $\delta_q \sim P$ varies linearly with pressure, for perturbations along the quartic mode given in Fig.~\ref{fig:quaritc} (a). 

Since the change in potential energy for perturbations along ``quartic" modes scales quadratically with the amplitude of the perturbation for $\delta \lesssim \delta_q$, it can be challenging to identify quartic modes. To count the number of quartic modes, we decompose the dynamical matrix into two components, the stiffness matrix $H$ and stress matrix $S$, where $M=H+S$~\cite{stiffness:DonevPRE2007,stiffness:SchreckPRE2012}. The stiffness matrix only depends on the geometry of the system (not the interaction potential or pressure), 
\begin{equation}
\label{eq:stiffness}
    H_{kl} = \sum_{i>j} \frac{\partial^2 V}{\partial (\vec{r}_{ij}/\sigma_{ij})^2} \frac{\partial(r_{ij}/\sigma_{ij})}{\partial {\vec r}_k} \frac{\partial(r_{ij}/\sigma_{ij})}{\partial {\vec r}_l}, 
\end{equation}
where $k$ and $l$ loop over all $N$ particle coordinates. Previous work has shown that quartic modes ${\vec \xi}^i$ in $M$ have non-zero eigenvalues $e^i$ at non-zero pressure; however, the same eigenmode yields $H {\vec \xi}^i = h^i {\vec \xi}^i$, where  $h^i=0$~\cite{stiffness:SchreckPRE2012}. Therefore, for each repulsive polymer packing, we calculate the number of quartic modes $N_q = H_0 - M_0$, where $M_0$ and $H_0$ are the number of zero modes in the dynamical matrix and stiffness matrix, respectively.  We find that packings of repulsive polymers are hypostatic at jamming onset with $N_c + N_w + N_b < N^{\rm iso}_c$, where $N_b$ is the number of polymer bonds. However, the number of missing contacts $N_m = N^{\rm iso}_c - N_c - N_w - N_b$ equals the number of quartic modes $N_m = N_q$ for each repulsive polymer packing. As shown Fig.~\ref{fig:U_vs_delta_N} (b), we find identical finite-size scaling and collapse of the excess number of contacts $\Delta N$ versus $V_r N^3$ for packings of repulsive polymers and packings of repulsive disks, where $\Delta N = N_{c} + N_{w} + N_{b} + N_{q} - (2(N-N_f) + f(d) + 1)$ for packings of repulsive polymers.

\subsection{Attractive disk and polymer packings are hyperstatic, but effectively isostatic}
\label{sec:results_iso}

Above, we showed that repulsive packings are isostatic at jamming onset and obey power-law scaling relations for $\omega^{\ast}$ and $\Delta N$ versus $V_r N^3$. In addition, we find that attractive monomer and polymer packings not only possess similar core packing fractions as their repulsive counterparts, but also follow the same power-law scaling relation for $\omega^{\ast}$ versus $V_r N^3$. Can attractive disk and polymer packings be viewed as effectively isostatic as well?

Typical contact counting analyses consider a constraint as the onset of any non-zero interaction between particles or between a particle and a wall. Thus, for attractive systems in Eq.~\ref{eq:ra}, a contact could be defined as an interparticle separation that satisfies $r_{ij} / \sigma_{ij} < 1 + \alpha$. With this definition, packings of attractive monomers and polymers are highly hyperstatic. However, previous studies have suggested that weak long-range attractions are relatively unimportant for determining the mechanical properties of attractive solids~\cite{sticky:XuPRL2007}. Remarkably, using the attractive potential in Eq.~\ref{eq:ra}, we find that if we count contacts as those with interparticle separations with $r_{ij}/r_{\beta} < 1$, packings of attractive monomers are effectively isostatic for small $V_r$, i.e. $N_c(r_{ij} < r_{\beta}) = N_c^{\rm iso}$, where $N^{\rm{iso}}_c = 2N - f(d)$ and $f(d)=3$ for the two uniform translations and a single rotation that have no energy cost for attractive packings with open boundary conditions. In Eq.~\ref{eq:ra}, $r_{\beta}$ indicates a change in the interaction stiffness. For $r_{ij}/\sigma_{ij} < r_\beta$, $|\partial^2 V / \partial r_{ij}^2| \sim \epsilon$, whereas $|\partial^2 V / \partial r_{ij}^2| \sim k/\epsilon \sim \beta$ tends to zero as $\beta \rightarrow 0$. In Fig.~\ref{fig:U_vs_delta_N} (c), we show that   $\Delta N = N_c(r_{ij} < r_{\beta}) - N_c^{\rm iso}$ obeys the same power-law scaling with $V_r N^3$ as found for packings of repulsive disks and polymers.

\begin{figure}
\begin{center}
\includegraphics[width=0.875\columnwidth]{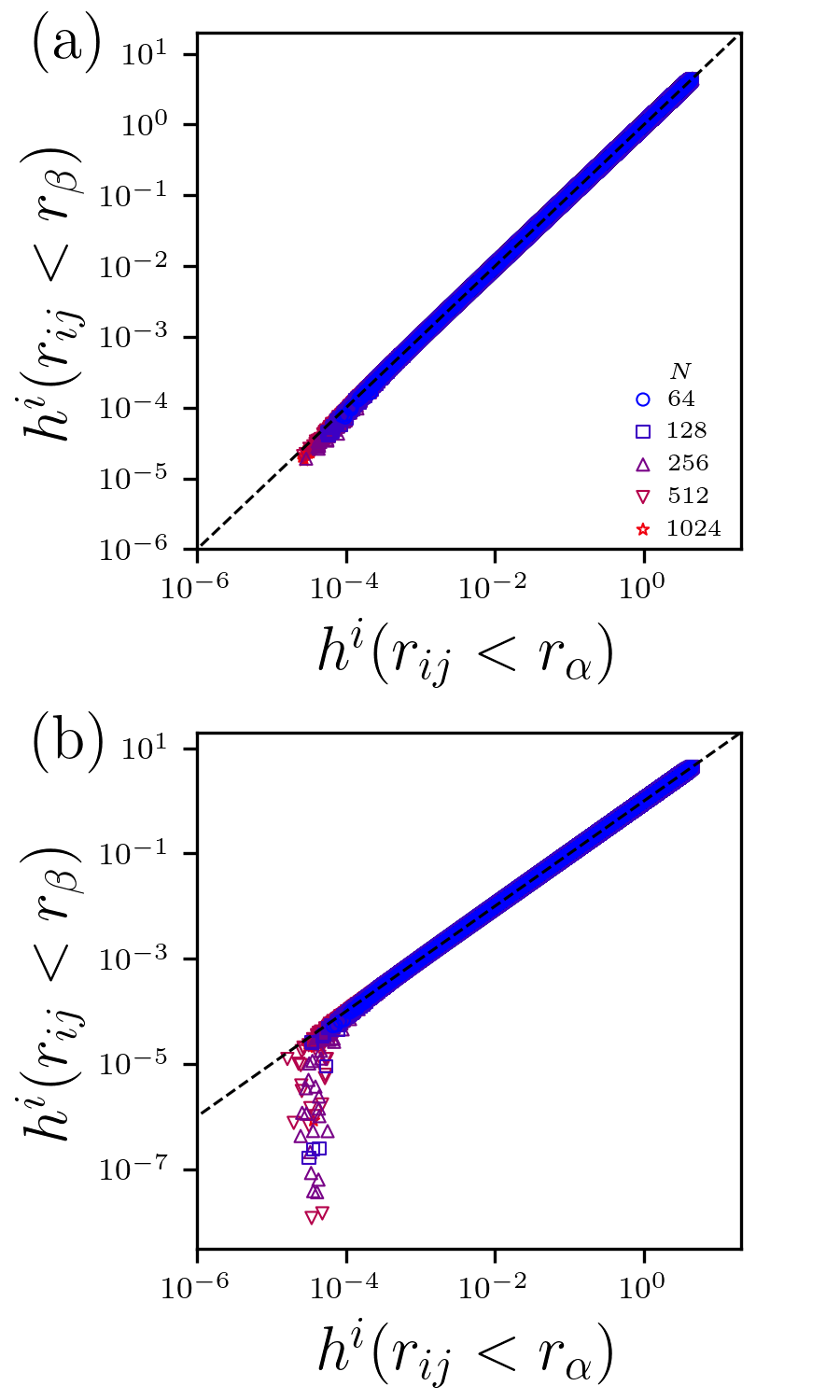}
\caption{The eigenvalues  $h^i(r_{ij}<r_{\beta})$ of the stiffness matrix $H(r_{ij} < r_{\beta})$ for attractive packings with contacts defined by $r_{ij} < r_{\beta}$ plotted versus the eigenvalues $h^i(r_{ij} < r_{\alpha})$ for $H(r_{ij} < r_{\alpha})$ with contacts defined using the full attractive potential for attractive (a) disks and (b) polymers as a function of system size, $N=64$ (circles), $128$ (squares), $256$ (upward triangles), $512$ (downward triangles), and $1024$ (stars) colored from blue to red with increasing system size. The black dashed line indicates $h^i(r_{ij}<r_{\beta})=h^i(r_{ij}<r_{\alpha})$.}
\label{fig:rb_eq}
\end{center}
\end{figure}

We have shown that if we define contacts for packings of attractive disks as those with $r_{ij} < r_{\beta}$, attractive disk packings are effectively isostatic (for $V_r N^3 \ll 1$) and $\Delta N$ versus $V_r N^3$ obeys similar power-law scaling as that found for isostatic repulsive packings.  However, do attractive packings with contacts defined by $r_{ij} < r_{\beta}$ possess any zero-energy modes?  To address this question, we construct the stiffness matrix from contacts defined by $r_{ij} /r_{\beta} < 1$ in attractive disk packings. We then calculate the stiffness matrix eigenvalues $h^i (r_{ij} < r_{\beta})$ and compare them to the eigenvalues of the stiffness matrix $h^i(r_{ij}<r_{\alpha})$ using contacts defined by the full attractive potential. We not only find that attractive disk packings with contact networks defined by $r_{ij} < r_{\beta}$ are effectively isostatic, but also that $H(r_{ij} < r_{\beta})$ has no non-trivial zero-energy modes, $h^i (r_{ij} < r_{\beta})>0$. We further show in Fig.~\ref{fig:rb_eq} (a) that for the attractive disks the eigenvalues $h^i (r_{ij} < r_{\beta})$ are nearly identical to the eigenvalues $h^i(r_{ij}<r_{\alpha})$.

\begin{figure}
\begin{center}
\includegraphics[width=0.875\columnwidth]{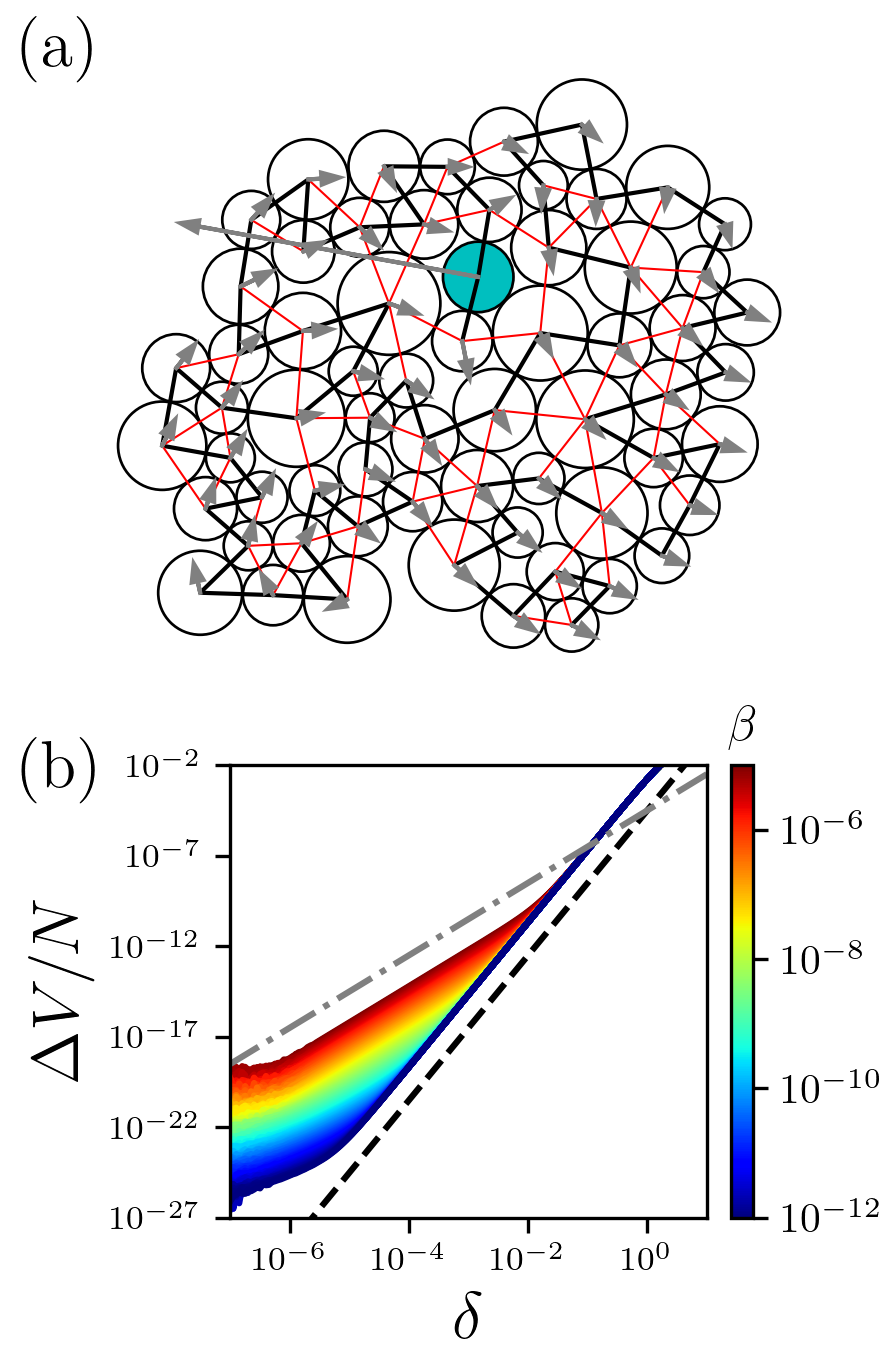}
\caption{(a) Illustration of an attractive polymer packing with $N=64$ and $\beta =10^{-5}$. We highlight the quartic mode in (b) with grey arrows.  The red lines indicate contacts that satisfy $r_{ij} < r_{\beta}$ and the black lines indicate the polymer backbone. $N_c(r_{ij}<r_{\beta}) + N_b = 124$ and $N^{\rm{iso}}_c=2N-3=125$ and therefore the packing is missing a single contact. The cyan-shaded particle has no non-bonded contacts with $r_{ij} < r_{\beta}$ and has the largest contribution to the quartic mode. (b) Change in the total potential energy $\Delta V/N$ following a perturbation with amplitude $\delta$ applied along the quartic mode of the dynamical matrix in (a) for increasing attractive strength $\beta$ (curves shaded from blue to red). The grey dot-dashed and black dashed lines indicate slopes of $2$ and $4$.}
\label{fig:scd_hypo}
\end{center}
\end{figure}

Are packings of attractive polymers effectively isostatic using the same definition of interparticle contacts as packings of attractive disks? When defining contacts as $r_{ij} / r_{\beta} < 1$, some attractive polymer packings appear to be hypostatic with $N_c(r_{ij} < r_{\beta}) + N_b < N_c^{\rm{iso}}$. For example, in Fig.~\ref{fig:scd_hypo} (a), we show an attractive polymer packing with $N_c(r_{ij}<r_{\beta}) + N_b = 124$ and $N^{\rm{iso}}_c=2N-3=125$ and therefore this packing is missing a single contact. We find that the lowest non-trivial eigenmode of the dynamical matrix $M$ is very similar to a quartic mode in a jammed repulsive polymer packing, where the largest contribution to the mode is perpendicular to a $\sim 180 \degree$ bond angle. For repulsive polymer packings, the number of quartic modes satisfies $N_q = H_0 - M_0$. In attractive polymer packings with missing contacts, $H_0 = M_0$ and $N_q$ appears to be $0$. However, we show in Fig.~\ref{fig:scd_hypo} (b) that when we perturb the attractive polymer packing in Fig.~\ref{fig:scd_hypo} (a) along the possible quartic mode of $M$, the change in the total potential energy $V=V^{anb}+V^{b}$ versus the perturbation amplitude $\delta$ scales as $\Delta V \sim \delta^4$ for $\delta > \delta_q \sim \beta^2$.

When we consider $H(r_{ij}<r_{\alpha})$ and $M(r_{ij} < r_{\alpha})$, we find that $N_q = H_0 - M_0 = 0$ even for attractive polymer packings that are hypostatic. However, we find that $H_0(r_{ij} < r_{\beta}) > H_0(r_{ij} < r_{\alpha})$ for attractive polymer packings with missing contacts. Therefore, for attractive polymer packings, we count the number of quartic modes $N_q$ as the number of non-trivial zero modes in $H(r_{ij} < r_{\beta})$. When including these $N_q$ quartic modes, we find that $\Delta N=N_c(r_{ij}<r_{\beta}) + N_b + N_q$ versus $V_r N^3$ obeys the same power-law scaling and finite-size collapse as packings of repulsive disks, repulsive polymers, and attractive disks. (See Fig.~\ref{fig:U_vs_delta_N} (d)). While packings of attractive polymers are effectively isostatic, we also find that the low-frequency eigenvalues of the stiffness matrix $h^i(r_{ij} < r_{\beta})$ deviate from those $h^i(r_{ij}<r_{\alpha})$ defined using the full attractive potential (Fig.~\ref{fig:rb_eq} (b)). This result indicates that quartic modes in attractive polymer packings are more sensitive (compared to the low-frequency stiffness matrix eigenvalues of packings of attractive disks) to the addition of the weak long-range attractions of the full attractive potential.

\begin{figure}
\begin{center}
\includegraphics[width=\columnwidth]{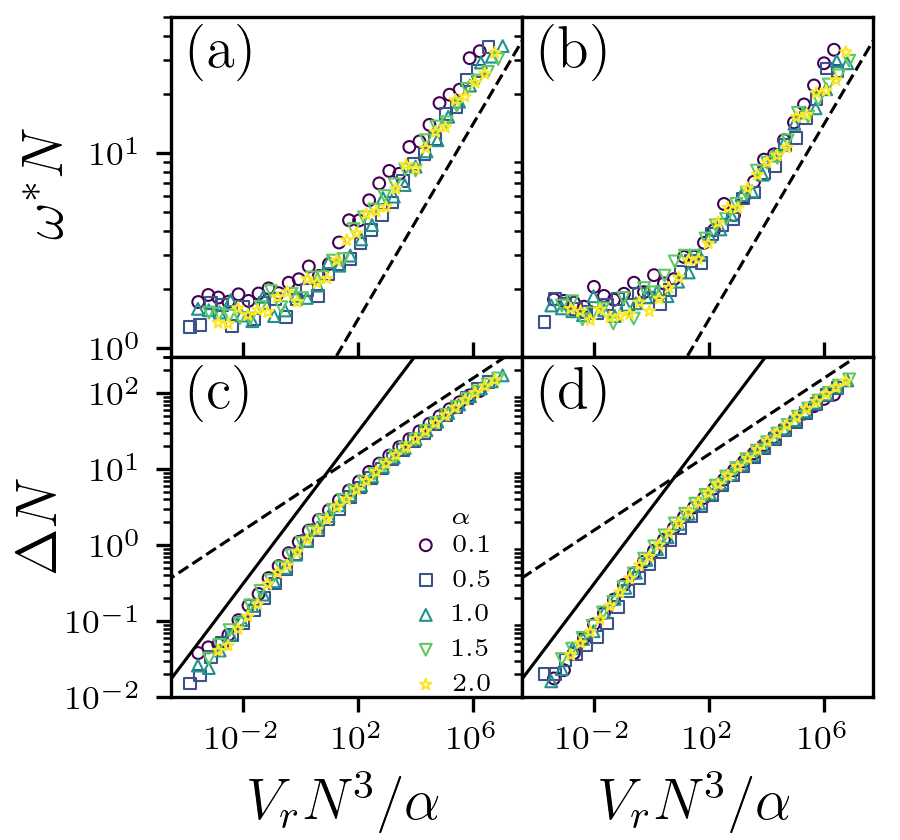}
\caption{Characteristic plateau frequency of the vibrational density of states $\omega^{\ast}$ plotted versus $V_r N^{3} / \alpha$ for attractive (a) disk and (b) polymer packings and the excess contacts $\Delta N$ plotted versus $V_r N^{3} / \alpha$  for attractive (c) disk $(\Delta N = N_c(r_{ij}<r_{\beta}) - N_c^{\rm{iso}})$ and (d) polymer $(\Delta N = N_c(r_{ij}<r_{\beta}) + N_b + N_q - N_c^{\rm{iso}})$ packings and with varying attractive ranges, $\alpha=0.1$ (circles), $0.5$ (squares), $1.0$ (upward triangles), $1.5$ (downward triangles), and $2.0$ (stars) colored purple to yellow with increasing $\alpha$ for $N=256$. In (a) and (b), the dashed lines indicate slopes of $0.25$ and in (c) and (d) the dashed and solid lines indicate slopes of $0.25$ and $0.5$ respectively.}
\label{fig:alpha_collapse}
\end{center}
\end{figure}

Are attractive disks and polymers still effectively isostatic when varying the range of the attractive interaction $\alpha$? We change the attractive range in small steps, $\alpha = \alpha_0 \pm \Delta \alpha$, where $\alpha_0=1.5$ and $\Delta \alpha = 0.01$ with each $\alpha$ increment followed by energy minimization. In Fig.~\ref{fig:alpha_collapse} (a) and (b), we show the scaling of $\omega^{\ast}N$ versus $V_r N^3 /\alpha$ for $0.1 \le \alpha \le 2$ for packings of attractive disks and polymers and find that $\omega^{\ast} N \sim (V_r N^3/\alpha)^{1/4}$ collapses the data for all values of $\alpha$. In Fig.~\ref{fig:alpha_collapse} (c) and (d), we show that packings of attractive disks and polymers are also effectively isostatic when defining contacts according to $r_{ij} < r_{\beta}$ for all $\alpha$. For all packings of attractive disks and polymers, $\Delta N >0$ and $\Delta N$ versus $V_r N^3/\alpha$ obeys the same scaling relation as that found for isostatic packings of repulsive disks and polymers. 

\section{\label{sec:conclusions}Conclusions and Future Directions}

In this work, we studied the connection between the collapse of attractive disk-shaped bead-spring polymers and the onset of jamming in packings of repulsive disks and polymers. This work was motivated by the fact that protein cores possess similar packing fractions to those of jammed packings of purely repulsive, disconnected amino-acid-shaped particles.  Is there a deep connection between attractive polymer collapse and compression-induced jamming or is the similarity fortuitous?

First, we showed that for packings of attractive disk-shaped bead-spring polymers to possess interior packing fractions similar to those in jammed repulsive disk packings, they must be quenched to temperatures much below the glass transition. To compare packings of attractive and repulsive disks and polymers, we developed a method to compress repulsive systems under open boundary conditions. We find that the average core packing fraction of repulsive disk and polymer packings under this protocol is similar to that generated by thermally quenching attractive disks and polymers. 

Previous studies have shown that repulsive disk packings at jamming onset are isostatic and possess an excess of low-frequency modes in the vibrational density of states, with a characteristic plateau frequency $\omega^{\ast} \sim \Delta N \sim (V_r N^3)^{1/4}$, where $\Delta N$ is the excess contact number, $\Delta N=N_c + N_w - N_c^{\rm iso}$, $V_r$ is the repulsive contribution to the potential energy, $N_c$ is the number of interparticle contacts, $N_w$ is the number of particle-wall contacts, and $N_c^{\rm{iso}}=2(N-N_r)+f(d)+1$. While repulsive polymer packings are typically hypostatic at jamming onset, the number of missing contacts equals the number of quartic modes $N_q$ and we find that repulsive polymers are effectively isostatic such that the excess contacts $\Delta N = N_c + N_w + N_b + N_q - N_c^{\rm{iso}}$ versus $V_r N^3$ obeys the same scaling form as that found for packings of repulsive disks, where $N_b$ is the number of polymer bonds and $N_c^{\rm iso}=2(N-N_f)+f(d)+1$. 

In overconstrained systems, the vibrational density of states $D(\omega) \rightarrow 0$ in the low-frequency limit~\cite{jamming:OHernPRE2003}. Here, we show that even though attractive disk and polymer packings are highly hyperstatic due to longer-range attractive interactions, they possess a plateau in the low-frequency region of $D(\omega)$ and that $\omega^{\ast} \sim (V_r N^3)^{1/4}$. Since this power-law scaling behavior for $\omega^{\ast}$ versus $V_r N^3$ is similar to that for packings of repulsive disks and polymers near jamming onset, it suggests that packings of attractive monomers and polymers with weak attractions are effectively isostatic. We find that if we define contacts as non-bonded pairs with $r_{ij} < r_{\beta}$, packings of attractive of monomers and polymers are effectively isostatic with $\Delta N = N_c(r_{ij}<r_{\beta}) + N_ q - N_c^{\rm{iso}} \sim (V_r N^3)^{1/4}$, where $N_c^{\rm{iso}}=2N-f(d)$. These results indicate that longer-range attractions provide an average compression force, but that the mechanical properties are controlled by the stronger short-range repulsive interactions. Note that scattering experiments on protein crystal structures have shown that they also possess a plateau in $D(\omega)$ at low frequencies, which suggests that proteins may also be viewed as effectively isostatic~\cite{glassyproteins:PerticaroliBPJ2014}.

Overall, we find that there is a deep connection between the interior packing fraction, low-frequency regions of the vibrational density of states, and isostaticity in all four systems: jammed packings of repulsive disks and polymers and thermally quenched, collapsed attractive disks and polymers. Note that we considered an interparticle potential with a discontinuous jump in its second derivative, and the location of the discontinuity corresponded to the definition of interparticle contacts that yields effective isostaticity. In future work, we will study interaction potentials where we can vary the magnitude of the change in the second derivative and the range over which it changes to understand the parameters that control whether attractive packings can be considered as effectively isostatic. 

Here, we established that for thermally quenched attractive disk-shaped bead-spring polymers to obtain interior packing fractions near values found for jammed packings of repulsive disks and polymers, they must be cooled below the glass transition temperature. Thus, the collapsed polymers we considered are glassy and the interior packing fraction can be increased by decreasing the cooling rate~\cite{aging:HuNatPhys2016}. Similarly, we have already shown that the packing fraction at jamming onset for packings of repulsive amino-acid-shaped particles spans the range $0.55 < \phi < 0.62$, where the average core packing fraction for protein x-ray crystal structures ($\langle \phi \rangle \sim 0.55$) is only obtained in the limit of rapid compression and energy minimization ~\cite{subgroup:MeiProteins2020}. In contrast, the current view of the protein energy landscape emphasizes that proteins fold in equilibrium to the global energy minimum~\cite{energylandscape:BryngelsonPNAS1987,energylandscape:LeopoldPNAS1992,energylandscape:OnuchicAnnRevPhysChem1997,energylandscape:PlotkinQRevBiophys2002}.

Our work suggests that experimentally determined protein cores can in principle reach packing fractions of $\phi = 0.62$ and yet, we find that they always possess the rapid thermal quench value of $\phi \sim 0.55$. In future work, we will generate packings using an all-atom hard-sphere model for proteins with stereochemical constraints (including constraints on the bond lengths, bond angles, and peptide bond dihedral angles $\omega$) using compression or thermal collapse with short-range attractive interactions, to verify that the cores in these model proteins can possess a range of packing fractions, $0.55 < \phi < 0.62$. These single protein packings will obey the geometric criteria of high-quality protein x-ray crystal structures (i.e. no non-bonded overlaps and bond lengths, bond angles, and backbone and side-chain dihedral angles will obey the statistics found for protein structures in the Protein Data Bank) and possess core packing fractions with $0.55 < \phi < 0.62$, but will not take on their native folds~\cite{subgroup:GrigasProSci2020,subgroup:GrigasProSci2022}. To investigate whether proteins in their native conformations can possess a range of core packing fractions, we will initialize these simulations with a given protein x-ray crystal structure, add short-range attractive, non-bonded atomic interactions with different strengths, thermally quench the system over a range of cooling rates, and measure the core packing fraction. Additionally, varying the attractive depth of the atomic interactions can be used to capture the range of hydrophobic interactions for different amino acids.

\begin{acknowledgments}
The authors acknowledge support from NIH Training Grant No. T32GM145452 and the High Performance Computing facilities operated by Yale’s Center for Research Computing.
\end{acknowledgments}

\appendix

\section{Generating repulsive disk and polymer packings in open boundary conditions}
\label{sec:jamming_appendix}

To generate static packings of repulsive disks and polymers under open boundary conditions, we apply an external central potential $V^{c}$ in Eq.~\ref{eq:comp} for all disks in the packing. With this central potential and in the limit of large damping paramaters, repulsive disk and polymer packings are highly disordered. However, with low damping parameters, thermal fluctuations can induce size segregation in packings of repulsive disks, with small disks slipping past large disks, which leaves only large disks on the surface and gives rise to crystallization. Therefore, we add a bias factor $\left( \sigma_i / \sigma_{\rm{max}} \right)^{\nu}$ to the compression force, such that larger disks feel larger compression forces. The exponent $\nu$ controls the strength of the bias factor.

\begin{figure}
\begin{center}
\includegraphics[width=0.875\columnwidth]{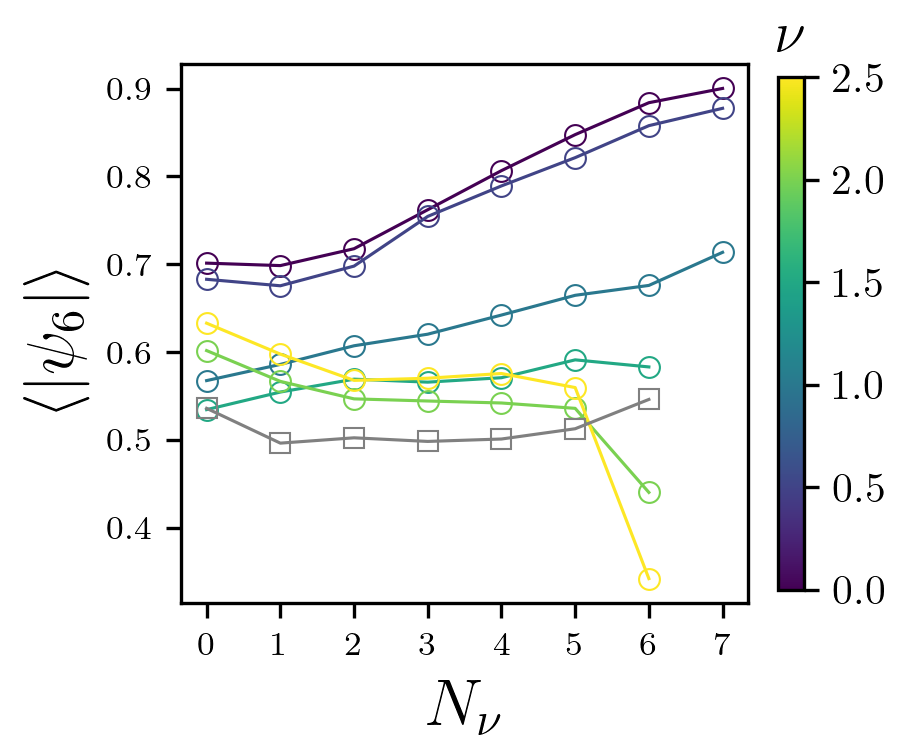}
\caption{The average hexatic order parameter $\langle \vert \psi_6 \vert \rangle$ plotted versus the number of Voronoi cells $N_{\nu}$ between each disk and the closest surface disk for varying exponents $\nu$ (increasing from purple to yellow) that control the strength of the bias factor of the compression force for packings of repulsive disks with $N=256$ prepared using $b=10^{-5}$.  As a comparison, we also show results for packings of attractive disks prepared at the same value of $b$ (grey squares).}
\label{fig:psi6_vs_nu}
\end{center}
\end{figure}

As shown in Fig.~\ref{fig:packing_vs_nv} (b), attractive disk and polymer packings do not size segregate and therefore we can calibrate the value of $\nu$ by comparing the structural properties of repulsive disk to those of attractive disk and polymer packings. In Fig.~\ref{fig:psi6_vs_nu}, we plot the average hexatic order parameter $\langle \vert \psi_6 \vert \rangle$ versus the number $N_{\nu}$ of Voronoi cells between a disk and the surface as a function of $\nu$ for packings of repulsive disks. As $\nu$ increases, the hexactic order decreases strongly for all values of $N_{\nu}$. However, the similarity between the repulsive and attractive disk packings decreases when $\nu \gtrsim 2.5$. Therefore, we use $\nu=2$ for preparing all repulsive disk packings in these studies.

\begin{figure}
\begin{center}
\includegraphics[width=0.875\columnwidth]{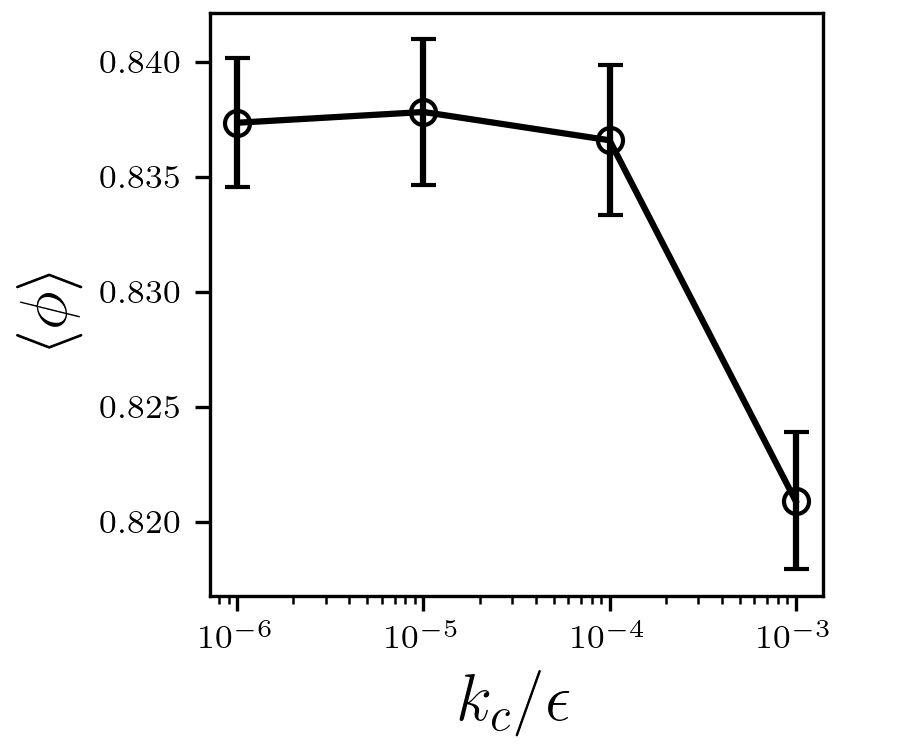}
\caption{The average core packing fraction of packings of repulsive disks plotted as a function of $k_c/\epsilon$ using $N=256$ and $b=0.1$.}
\label{fig:phi_vs_f0}
\end{center}
\end{figure}

We also studied the influence of the spring constant $k_c/\epsilon$ on the core packing fraction in packings of repulsive disks. The spring constant $k_c$ controls the effective rate of compression, which is known to influence the structural properties of jammed packings~\cite{jamming:OHernPRE2003}. In Fig.~\ref{fig:phi_vs_f0}, we plot the average core packing fraction $\langle \phi \rangle$ for $100$ repulsive disk packings for $N=256$ and $b=0.1$ versus $k_c/\epsilon$. When compressing with large $k_c/\epsilon$, the repulsive disk packings tend to be less densely packed and the packing fraction reaches a plateau for $k_c/\epsilon \lesssim 10^{-4}$. Therefore, we selected $k_c/\epsilon=10^{-4}$ to generate all repulsive disk packings.

\section{Identification of core disks}
\label{sec:rsasa_appendix}

To examine the packing fraction of the interior of disk and polymer packings in open boundaries, we must first quantitatively define which disks are considered as ``core'' versus ``non-core''. Here, we implement an algorithm first proposed by Lee and Richards~\cite{rsasa:LeeJMB1971} that is frequently used to measure the solvent-accessible surface area in proteins. In the case of disk and polymer packings in open boundaries, we place a probe disk of diameter $\sigma_p$ on the ``anchor'' disk that is furthest from the center of mass of the packing. We rotate the probe around the anchor disk in angle increments of $\Delta \theta = 0.1$ radians and check for overlaps with neighboring disks. If a new contact is made with the probe disk, the new contacting disk becomes the anchor disk. This process is repeated until the probe disk returns to the initial anchor disk. In proteins, $\sigma_p$ is given by the size of a water molecule so that the surface area swept out by the probe reflects the solvent-accessible surface area.

\begin{figure}
\begin{center}
\includegraphics[width=0.875\columnwidth]{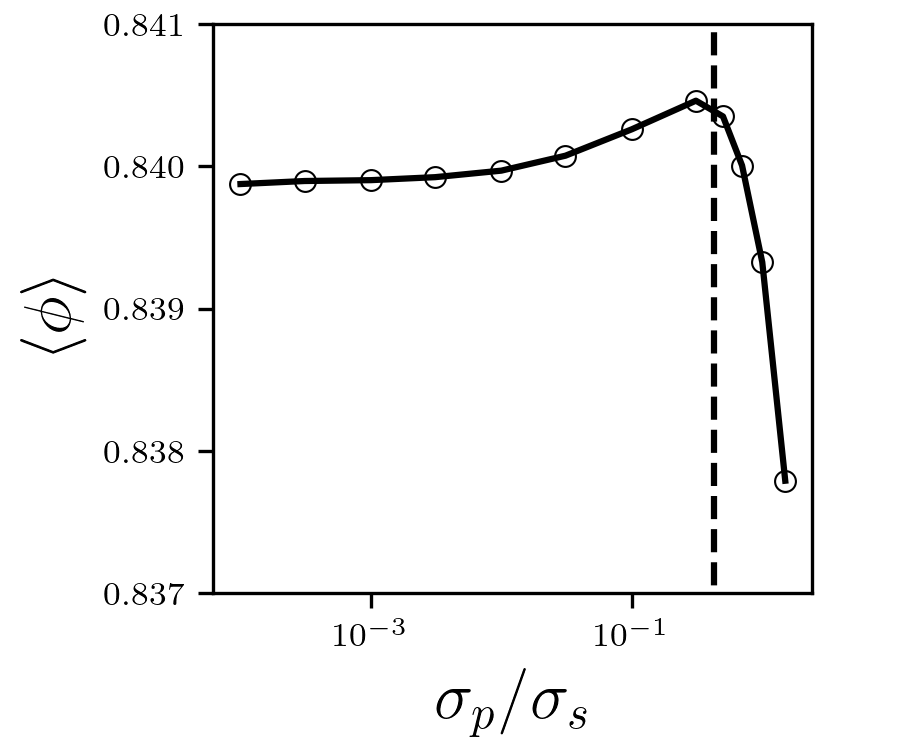}
\caption{The average core packing fraction $\langle \phi \rangle$ plotted versus the ratio of the surface probe diameter to the smallest disk diameter $\sigma_p / \sigma_s$ for packings of attractive polymers with $N=256$, $b=10^{-5}$, and $T_0/T_m=0.27$. The vertical dashed line indicates $\sigma_p/\sigma_s \sim 0.43$, which is the ratio of the diameter of a water molecule to an Alanine residue.}
\label{fig:phi_vs_sigmap}
\end{center}
\end{figure}

The size of the probe will determine which disks are labeled as core and thus affect the average core packing fraction $\langle \phi \rangle$. In Fig.~\ref{fig:phi_vs_sigmap}, we plot $\langle \phi \rangle$ versus $\sigma_p$ for $N=256$ attractive polymer packings  For large probe sizes, similar in size to the largest disk in the system, the core packing fraction decreases significantly as more surface-like (non-core) particles are included in the average. The core packing fraction plateaus as $\sigma_p \lesssim 0.4$. The typical probe size used to study proteins is the diameter of a water molecule $\sigma_p \sim 2.8~\text{\AA}$, whereas the maximum diameter of an Alanine residue is $6.6~\text{\AA}$, which yields the ratio, $\sigma_p / \sigma_s \sim 0.43$. In the studies in the main text, we chose a similar ratio $\sigma_p/\sigma_s = 0.1$.

\newpage

\end{document}